\documentclass[12pt]{article}
\usepackage[utf8]{inputenc}
\usepackage{amsmath}
\usepackage{amsfonts}
\usepackage{authblk}
\allowdisplaybreaks 
\usepackage{hyperref}
\usepackage{lmodern}
\usepackage[english]{babel}

\newcommand{\CS}{{\mathcal{C}}}
\newcommand{\G}[1]{\Gamma\left(#1 \right)}
\newcommand{\mFm}[4]{\, _{#1}F_{#1}\!\left( \genfrac{}{}{0pt}{}{#2}{ #3  }  \bigg| #4 \right) }
\newcommand{\pFq}[5]{\, _{#1}F_{#2}\!\left( \genfrac{}{}{0pt}{}{#3}{ #4  }  \bigg| #5 \right) }
\newcommand{\Poch}[2]{\left( #1 \right)_{#2}}
\newcommand{\Od}[1]{\mathcal{O}_{#1}}
\newcommand{\md}{\mathrm{d}}
\newcommand{\pd}[2]{\frac{\partial #1}{\partial #2}}

\newcommand{\suml}[2]{\sum\limits_{#1}^{#2}}
\newcommand{\intl}[2]{\int\limits_{#1}^{#2}}

\newcommand{\dtwo  }{\frac{d}{2} }
\newcommand{\twod  }{\frac{2}{d}  }
\newcommand{\dodd}{{d_{\text{odd}}}}
\newcommand{\deve}{{d_{\text{even}}}}
\newcommand{\bigspace}{\ \ \ \ \ \ \ \ \ \ \ \ \ \ \ }
\newcommand{\bigspaced}{\bigspace \bigspace }

\providecommand{\doi}[1]{\href{http://dx.doi.org/#1}{#1}}

\title{A closed form expression for the causal set d'Alembertian}

\author{Lisa Glaser${}^1$}

\affil{The Niels Bohr Institute, Copenhagen University, \\ Blegdamsvej 17, DK-2100 Copenhagen \O, Denmark\\ Email: glaser@nbi.dk}
\date{\today}

\begin{document}
\maketitle

\begin{abstract}
Recently a definition for a Lorentz invariant operator approximating the d'Alembertian in $d$-dimensional causal set space-times has been proposed.
This operator contains several dimension-dependent constants which have been determined for $d=2,\dots,7$. In this note we derive closed form expressions for these constants, which are valid in all dimensions.
Using these we prove that the causal set action in any dimension can be defined through this discrete d'Alembertian, with a dimension independent prefactor.
\end{abstract}

The causal structure of a causal Lorentzian manifold is a partial order on the set of events that comprise the manifold.
A remarkable result about Lorentzian geometries is that the causal structure fixes the space-time up to conformal transformations\cite{Hawking:1974,Malament:1977}.

Causal set theory is a proposal for quantum gravity that uses this insight to describe space-time.
In causal set theory one discretises space-time so that a minimal volume is associated to every space-time event.
The space-time can then be described as a locally-finite, partially-ordered set\cite{PhysRevLett.59.521}.
The partial order relation $\prec$ does then correspond to two elements being causally connected. 
Two events $x, y$ are related if their separation is time-like, and $x \prec y$ denotes that $x$ is to the past of $y$.
For a good introduction or a recent review on causal set theory see \cite{valdivia,Surya:2011yh}.

There a different ways of discretising a space-time. 
In a regular hyper-cubic lattice the discretisation is obtained using integer multiples of the discreteness scale $l$.
In causal set theory these coordinates are chosen randomly, according to a Poisson distribution.
The probability to find $m$ elements in a volume $V$  is
\begin{equation}\label{eq:poisson}
 P(m,V)=\frac{(\rho V)^m}{m!} \; e^{-\rho V}.
\end{equation}
The density $\rho$ can be expressed in terms of the discreteness scale  $l$ as $\rho=l^{-d}$. 
Using this procedure to generate a causal set from a given manifold is referred to as a ``sprinkling''.  

It has been argued \cite{Sorkin:2007qi} that a discrete Lorentzian manifold has to give up on locality.
The nearest neighbors of an event in a Lorentzian manifold, are all events light-like to it, namely its light-cone. 
Discretising the causal structure creates a graph of infinite valency, which leads to non-local effects. 
Conversely, a finite valency graph will break Lorentz invariance. 
This has been shown, assuming a random map that preserves the number volume correspondence is used to generate the graph, in \cite{Bombelli:2006nm}. There it is also shown that a causal set generated by a Poisson process does respect Lorentz invariance.
Therefore in discretising a Lorentzian manifold one has to choose between Lorentz invariance and locality.

Causal set theory keeps Lorentz invariance, thus sacrificing locality. 
Without locality the dynamics of the theory cannot follow the usual patterns. 
For example typically a derivative is discretized as a sum over nearest neighbors (cf. \cite{Calcagni:2012cv}). 
On an infinite valency graph such a sum would necessarily diverge, thus making the usual approach unsuitable for causal sets. 

This has led Sorkin to propose the operator $B^{(2)}$, which for a finite sprinkling density $\rho$ approximates the d'Alembertian operator acting on a scalar field in $2$ dimensions \cite{Sorkin:2007qi}
\begin{equation}
B^{(2)}\phi(x):=\frac{1}{l^2}\Big[- 2\phi(x)
+4\Big(\sum_{y \in L_1(x)}\!\!\!\phi(y)-
2\!\!\!\sum_{y\in L_2(x)}\!\!\!\phi(y)+\sum_{y\in L_3(x)}\!\!\!\phi(y)\Big)\Big]\,. \\
\label{Bop2}
\end{equation}
Here $y \in L_i$ indicates a sum over all elements in  the $i$-th layer to the past of the element $x$. The $i$-th layer $L_i(x)$ is the set of all elements $y$ for which $y \prec x$ and $n(x,y)=i-1$, where $n(x,y)$ is the number of elements that lie causally in between $x$ and $y$.
This definition is manifestly Lorentz invariant, since it only depends on the causal structure.

This operator was subsequently generalized to $4$ \cite{Benincasa:2010ac} and $d$ dimensions\cite{Dowker:2013vba}. 
In $d$ dimensions the operator $B^{(d)}$ is defined as
\begin{equation}
B^{(d)}\phi(x)=\frac{1}{l^2}\left( \alpha_d \phi(x) +\beta_d \sum\limits_{i=1}^{n}  C^{(d)}_i \sum\limits_{y \in L_{i}}\phi(y) \right) \, .
\end{equation}
Here we sum over $n$ layers of the causal set.
The cancellations between the contributions from the different layers $L_i$ require a minimal number of layers $n_{d}$ which is dimension dependent. 
We show here  that $n_d=\dtwo +2$ for even $d$ and $\frac{d-1}{2}+2$ for odd $d$, as chosen in \cite{Dowker:2013vba}, is indeed the minimal number required for the cancellation.
In \cite{Dowker:2013vba} we succeeded in determining the $C^{(d)}_i$ and the factors $\alpha_d$ and $\beta_d$ explicitly for dimensions $d=2, \dots , 7$.

To calculate this we used the fact that in a causal set that is well approximated by $d$ dimensional Minkowski space the average number of elements in the $i$-th layer can be calculated by integrating the probability density \eqref{eq:poisson},
\begin{equation}
I^{(d)}_i(l):=\int\limits_{J^-(x)} \mathrm{d}V \frac{(\rho V_d(x,y))^i}{i!} \,e^{-V_{d}(x,y)l^{-d}}.
\end{equation}
The integration runs over the causal past of the event $x$, and is easiest to solve in light-cone coordinates $u=\frac{1}{\sqrt{2}}(t-r)$ and $ v=\frac{1}{\sqrt{2}}(t+r)$.  
The volume $V_{d}(x,y)$ is the Alexandrov interval between $x$ and $y$, the continuum equivalent to the interval $n(x,y)$ in a causal set.
In flat space-time this expression can only depend on the proper time between the events $\tau_{(x-y)}= \sqrt{2 (u_x-u_y)(v_x-v_y)}$,
\begin{equation} \label{eq:defc}
V_{0\,d}(x,y) = c_d ((u_x-u_y)(v_x-v_y))^{d/2} = S_{d-2} \frac{1}{d(d-1) 2^{\dtwo-1}} ((u_x-u_y)(v_x-v_y))^{\dtwo  },
\end{equation}
where we introduce $c_d$ as a shorthand, and $S_{d-2}$ is the volume of the $d-2$ dimensional unit sphere.
The $i$-th interval $I^{(d)}_i(l)$ can be generated by acting $i$ times with $H= -l \pd{}{l}$ on $e^{-V_{d}(x,y)l^{-d}}$ and then correcting by a constant to obtain the factor $1/i!$.
Any sum over layers 
can then be generated by applying a power series in $H$.
We define the operator $\Od{d}$ as the power series in $H$ that generates the sum over past neighborhoods for the d'Alembertian.
\begin{equation}
\Od{d} \int\limits_{J^-(x)} \mathrm{d}V \exp(-l^{-d} V) = \sum_{i =1}^{n_d} C^{(d)}_i  \int\limits_{J^-(x)} \mathrm{d}V \frac{(l^{-d} V)^{i-1}}{(i-1)!} \exp(-l^{-d} V) \;.
\end{equation} 
We found that for even dimensions $d=2n$,  
\begin{eqnarray}\label{eq:op}
\Od{2n}=\frac{(H+2)(H+4) \dots (H+2n+2)}{2^{n+1}(n+1)!}\,,
\end{eqnarray}
and for odd dimensions, $d = 2n +1$, $\Od{2n+1} = \Od{2n}$. 

Using this operator $\Od{d}$ the average operator $\bar{B}^{(d)}$  on a causal set that is well approximated by $d$ dimensional Minkowski space can be written as
\begin{equation}\label{eqn:curved}
\bar{B}^{(d)} \phi(x) = \alpha_d{l^{-2}}  \phi(x)+ {\beta_d}{l^{-(d+2)}}  
\Od{d} \!\! \! \int\limits_{J^-(x)} \!\! \mathrm{d}V \phi(y) e^{-V_d(x,y)l^{-d}}\,. 
\end{equation}
In the infinite density limit this reproduces the d'Alembertian exactly. 
\begin{equation}
\lim_{l \rightarrow 0} \bar{B}^{(d)}\phi(x)
= \Box^{(d)} \phi(x)\,.
\end{equation}
For a given number of layers $n_d$, this equation fixes the constants $\alpha_d$, $\beta_d$ and $C_i^{(d)}$. 
The constants $\alpha_d$ and $\beta_d$ are required to satisfy 
\begin{align}
	\frac{1}{\beta_{d}}&= \lim_{l \rightarrow 0} \frac{S_{d-2} }{2(d-1)l^{d+2}} \Od{d} \intl{-\infty}{0} \md u \intl{u}{0} \md v  \left( \frac{v-u}{\sqrt{2}} \right)^{d} e^{-l^{-d} V_{0\,d}(u,v) } \\
\frac{\alpha_d }{\beta_d}&= - \lim_{l \rightarrow 0} 
\frac{ S_{d-2}}{l^{d}}  \Od{d} \intl{-\infty}{0} \mathrm{d}u \intl{u}{0}  \mathrm{d}v 
\left(\frac{v-u}{\sqrt{2}}\right)^{d-2} 
e^{-l^{-d} V_{0\,d}(u,v)} \;.
\end{align}
for any dimension. In \cite{Glaser:2013pca} integrals of this type were solved to a closed form expression and we now use this to find $\alpha_d$ and $\beta_d$.

In section \ref{sec:ab} of this note we will calculate this closed form expression and find that
\begin{align}
\beta_d&= \begin{cases} 
\frac{2 \; \G{\frac{d}{2}+2} \G{\frac{d}{2}+1}}{\G{\frac{2}{d}} \G{d}} \; c_d^{\frac{2}{d}} & \text{for even }d\\
  \frac{d+1}{2^{d-1}\G{\frac{2}{d}+1}} \; c_d^\frac{2}{d} & \text{for odd }d \\ 
 \end{cases}\\
	\intertext{and}
	\alpha_d&= \begin{cases}
\frac{- 2 c_d^{\frac{2}{d}}}{\G{\frac{d+2}{d}}} & \text{for even }d \\
	\frac{- c_d^{\frac{2}{d}}}{\G{\frac{d+2}{d}}} & \text{for odd }d \;.\\ \end{cases}
\end{align}
Using the same techniques we can also define generating functions for the the coefficients $C_i$ and obtain a simple form
\begin{align}
C_i^{(\deve)}=\sum_{k=0}^{i-1} \binom{i-1}{k} (-1)^k \, \frac{\G{\dtwo(k+1)+2}}{\G{\dtwo+2}{\G{1+\frac{d k}{2}}}}\\
C_i^{(\dodd)}=\sum_{k=0}^{i-1} \binom{i-1}{k} (-1)^k \, \frac{\G{\dtwo(k+1)+\frac{3}{2}}}{\G{\frac{d+3}{2}}{\G{1+\frac{d k}{2}}}} \;.
\end{align}

In \cite{Benincasa:2010ac} it was observed that the d'Alembertian acting on a scalar field in a curved space-time will obtain corrections proportional to the Ricci scalar of that manifold.
\begin{align}
\lim_{l \rightarrow 0} 
{\bar{B}}^{(d)}\phi(x) &= \Box^{(d)} \phi(x) + a_d R(x)\phi(x) 
\end{align}
Applying this operator to a constant field gives a measure of the Ricci scalar of that manifold at a given event, summing over all events of the manifold should lead to the analog of the Einstein Hilbert action.
This leads to a proposal for a causal set action
\begin{equation}
\frac{1}{\hbar}\mathcal{S}^{(d)}(\CS) = \zeta_d \bigg[ N + \frac{\beta_d}{\alpha_d} \sum_{i = 1}^{n_d} C_i ^{(d)} N_i \bigg]\,,
\end{equation}
where $N$ is the size of the causal set, $N_i$ the number of intervals of size $n(x,y)=i-1$ in the entire causal set, $\zeta_d = - \alpha_d \left(\frac{l}{l_p}\right)^{{d-2}}$ and $l_p^{d-2} = 8 \pi G\hbar$.

The case $d=2$  has been studied for flat regions of Minkowski space-time, a cylinder spacetime and the topology  changing trousers \cite{Benincasa:2010as}.
The action can also be used for Monte Carlo simulations \cite{Surya:2011du}.
The factor $a_d$ has been conjectured to be $-\frac{1}{2}$ for all dimensions, which we will prove in section \ref{sec:r}.

In the first section we calculate the closed form expression for $\alpha_d$ and $\beta_d$. In the second section we present some attempts to find a simple expression for the coefficients $C^{(d)}_i$ and in the last section we prove that the coefficient relating the operator $B^{(d)}$ to the Ricci scalar $R$ is $-\frac{1}{2}$ for all dimensions.

\section{\label{sec:ab}Finding \texorpdfstring{$\alpha_d$}{alphad} and \texorpdfstring{$\beta_d$}{betad} in general dimensions}

As mentioned above we can define $\beta_d$ and $\frac{\alpha_d}{\beta_d}$ through the integrals
\begin{align}
	\frac{1}{\beta_{d}}&= \lim_{l \rightarrow 0} \frac{S_{d-2} }{2(d-1)l^{d+2}} \Od{d} \intl{-L}{0} \md u \intl{u}{0} \md v  \left( \frac{v-u}{\sqrt{2}} \right)^{d} e^{-l^{-d} V_{0\,d}(u,v) } \label{eq:defbeta} \\
\frac{\alpha_d }{\beta_d}&= - \lim_{l \rightarrow 0} 
\frac{ S_{d-2}}{l^{d}}  \Od{d} \int_{-L}^0 \mathrm{d}u \int_u^0  \mathrm{d}v 
\left(\frac{v-u}{\sqrt{2}}\right)^{d-2} 
e^{-l^{-d} V_{0\,d}(u,v)} \label{eq:defalphabeta} \;.
\end{align}
In \cite{Dowker:2013vba} these integrals are solved for $L=\infty$, which corresponds to a past infinite causal set. Here, for technical reasons we introduce a cut-off $L$. Since $\alpha_d$ and $\beta_d$ are dimensionless quantities they can only depend on the ratio $L/l$. Thus taking the limit $l\to 0$ is equivalent to taking the limit $L\to \infty$ and the final expressions will be independent of $L$.  

These integrals can be solved by expanding the exponential as power series and expressing the $(v-u)$ terms through binomial sums.
After that the operator $\Od{d}$ has to be applied, at which point the discussion splits into odd and even parts. 
Once the closed form expressions are found we can use hypergeometric function identities to take the limit $l\to 0$ and thus find the desired expressions.
We will start this endeavor by calculating $\beta_d$.
\subsection{A general calculation for \texorpdfstring{$\beta_d$}{betad}}
In this subsection we will give an outline of how we calculate $\beta_d$, a much more detailed account of which can be found in appendix \ref{app:beta}.
To solve the integrals in $u$ and $v$ we do a power series expansion, after which the actual integration becomes simple
\begin{align}
&\intl{-L}{0} \md u \intl{u}{0} \md v  \left( \frac{v-u}{\sqrt{2}} \right)^{d} e^{-l^{-d}  c_d (uv)^{\dtwo  }} \\
&=\suml{k=0}{d}\binom{d}{k} \frac{(-1)^k}{2^{\dtwo  }} \suml{n=0}{\infty} \frac{\left(-l^{-d}c_d \right)^n}{n!} \frac{L^{d(n+1)+2}}{(\dtwo  n+k+1)(d(n+1)+2)}\;. \label{eq:intbeta1}
\end{align}
Next we apply the operator $\Od{d}$. At this point we will split the discussion into odd and even dimensions.

Applying the operator 
for even dimensions we find: 
\begin{align}
&\frac{\Od{\deve}}{l^{d+2}} \intl{-L}{0} \md u \intl{u}{0} \md v  \left( \frac{v-u}{\sqrt{2}} \right)^{d} e^{-l^{-d} c_d (uv)^{\dtwo  }}  \\
=&\suml{k=0}{d}\binom{d}{k} \frac{(-1)^k }{2^{\dtwo} (k+1)(d+2) c^{1+\twod}} z^{1+\twod}   \mFm{\dtwo+1}{\twod(k+1),\twod +1,\dots, 2 }{\twod(k+1)+1,\twod , \dots, 1 }{-z} \;.\label{eq:sumeve}
\end{align}
We introduced  $z= c_d (\frac{L}{l})^d$, a shorthand that we will use repeatedly throughout the paper.
The $\twod+1,\dots,2$ in the upper row stand for $\frac{d}{2}$ terms in the sequence $\twod j +1$ with $j=1,\dots,\frac{d}{2}$ (in the lower row $\twod,\dots,1$ stand for the sequence  $\twod j$ with $j=1,\dots,\frac{d}{2}$). 
We will in general use this shorthand for hypergeometric functions, the first and last term of the sequence will always be given and the steps in between be equidistant. 

The limit $l \to 0$ is equivalent to the limit $z \to \infty$. To calculate this limit we will use the following identities, 
\begin{align}
&\lim_{z\rightarrow \infty} e^{z} \mFm{q}{a_{1},\dots, a_{q}}{a_{1}-1,\dots, a_{q}-1}{-z}  = \frac{(-z)^{q}}{\prod\limits_{j=1}^{q} (a_{j}-1)} +O\left( (-z)^{q-1}\right)  \label{eq:limit0} \\
&	\lim_{z\rightarrow \infty} z^{a_{0}} \mFm{q+1}{a_{0},a_{1}, \dots,a_{q}}{a_{0}+1,a_{1}-1, \dots,a_{q}-1}{-z} = \G{a_{0}+1}  \prod\limits_{j=1}^{q}\frac{a_{j}- a_{0} -1}{a_{j}-1} \label{eq:limit}
\end{align}
which are proven in appendix \ref{app:limitident}.
In even dimensions when taking the limit $z\to \infty$ the terms in the sum in equation \eqref{eq:sumeve} split into three categories depending on $k$.
For terms where $k < \dtwo$ the first term in the hypergeometric function can be simplified against one of the later terms, leaving us with a function of the form \eqref{eq:limit0}. Thus the terms for $k<\dtwo$ do not contribute. If $k\geq \dtwo$ no simplification occurs and the hypergeometric function is of the form \eqref{eq:limit} with leading order term $z^{-\twod(k+1)}$. For $k>\dtwo$ this falls of faster than $z^{1+\twod}$, therefore these terms do not contribute. The only finite contribution in the limit comes from the term $k=\dtwo$.
\begin{align}
\lim_{z\to \infty} &\binom{d}{\dtwo} \frac{(-1)^\dtwo}{2^{\dtwo} (\dtwo+1)(d+2) c^{1+\twod}} z^{1+\twod} \mFm{\dtwo+1}{\twod+1, \twod +1,\dots, 2 }{\twod+2,\twod , \dots, 1 }{-z}\\
=&\binom{d}{\dtwo} \frac{1}{2^{\dtwo-1}d (d+2) c^{1+\twod}} \G{\frac{2+d}{d}} 
\end{align}
We can insert this into equation \eqref{eq:defbeta} and find
\begin{align}
	\beta_\deve &=\frac{2 \; \G{\frac{d}{2}+2} \G{\frac{d}{2}+1}}{\G{\frac{2}{d}} \G{d}} \; c_d^{\twod} \label{eq:solbetaeven}
\end{align}
Next we calculate $\beta_\dodd$, by applying the operator for odd dimensions to equation \eqref{eq:intbeta1}
\begin{align}
&\frac{1}{l^{d+2}}\Od{\dodd} \intl{-L}{0} \md u \intl{u}{0} \md v  \left( \frac{v-u}{\sqrt{2}} \right)^{d} e^{-l^{-d} c_d (uv)^{\dtwo  }}  \\
=& \suml{k=0}{d}\binom{d}{k} \frac{(-1)^k}{2^{\dtwo} \; l^{d+2}} \frac{L^{2+d}}{ (k+1)(d+2)} \nonumber \\
& \bigspace \mFm{\frac{d+1}{2}+2}{1+\twod,\twod (k+1), \twod+1, \dots, \frac{1}{d}+2}{2+\twod,\twod (k+1)+1, \twod, \dots, \frac{1}{d}+1}{- z }\;. \label{eq:oddintermediate}
\end{align}
We can not immediately use the identities \eqref{eq:limit0} and \eqref{eq:limit} on this, since the sequences in the argument do not have the required form. We can however expand the hypergeometric function \cite{07.31.03.0008.01}\footnote{Please refer to appendix \ref{app:subsubbodd} for more detail.}.
This leads to
\begin{align}
&\lim_{z\rightarrow \infty} z^{1+\twod} \mFm{\frac{d+1}{2}+2}{1+\twod,\twod (k+1), \twod+1,  \dots, \frac{1}{d}+2}{2+\twod,\twod (k+1)+1, \twod, \dots, \frac{1}{d}+1}{-z}\\
&=\lim_{z\rightarrow \infty} \frac{z^{1+\twod}}{1-\twod k} \bigg\{
 (1+\twod) \mFm{\frac{d+1}{2}+1}{\twod (k+1), \twod+1,  \dots, \frac{1}{d}+2}{\twod (k+1)+1, \twod,  \dots, \frac{1}{d}+1}{-z}  \nonumber \\
& \bigspace  - \twod (k+1) \mFm{\frac{d+1}{2}+1}{1+\twod , \twod+1,  \dots, \frac{1}{d}+2}{2+\twod , \twod,  \dots, \frac{1}{d}+1}{-z}  \bigg\}\;.
\end{align}
Comparing to equations \eqref{eq:limit0} and \eqref{eq:limit} we can see that the first term does not contribute.  
For $\twod (k+1)< 1+\frac{1}{d}$, the term $\twod(k+1)$ can be simplified against one of the later terms, and thus the hypergeometric function goes to zero as $z \to \infty$ by equation \eqref{eq:limit0}. For $k>\dtwo$ ,the leading order behavior can be determined as $z^{-\twod(k+1)}$ using \eqref{eq:limit}. This goes to zero faster than $z^{1+\twod}$, and thus also vanishes in the limit.
This leaves us with the second term
\begin{align}
 -\lim_{z\rightarrow \infty} \frac{\twod (k+1)}{1-\twod k} z^{1+\twod}  \mFm{\frac{d+1}{2}+1}{1+\twod , \twod+1, \dots, \frac{1}{d}+2}{2+\twod , \twod,  \dots, \frac{1}{d}+1}{-z}  \;.
\end{align}
We can use equation \eqref{eq:limit} to take the limit $z \to \infty$, and insert the result into the sum in equation \eqref{eq:oddintermediate} above which we can then solve
\begin{align}
=& \suml{k=0}{d}\binom{d}{k} \frac{(-1)^{\frac{d+1}{2}+1+k}}{2^{\dtwo} (\dtwo-k)} \frac{l^{2+d} z^{1+\dtwo} }{ (d+2)c_d^{1+\dtwo}}   \frac{\G{2+\twod} \G{1+\dtwo}}{\G{\frac{d+1}{2}+1} \sqrt{\pi}}  \\
=& \frac{2^{1+\dtwo} \G{\frac{2}{d}+1}}{d(d+1)}\frac{l^{2+d}  }{c_d^{1+\dtwo}} \;.\label{eq:oddlimitfin}
\end{align}
Inserting this into the definition of $\beta_d$ above \eqref{eq:defbeta} leads to
\begin{align}
\beta_{d_{\text{odd}}}= \frac{d+1}{2^{d-1}\G{\frac{2}{d}+1}} \; c_d^\twod \label{eq:solbetaodd} \;.
\end{align}

Thus we have calculated the previously unknown closed form expression for $\beta_d$ in odd and even dimensions

\begin{align}
\beta_d= \begin{cases} 
\frac{2 \; \G{\frac{d}{2}+2} \G{\frac{d}{2}+1}}{\G{\frac{2}{d}} \G{d}} \; c_d^{\frac{2}{d}} & \text{for even }d\\
  \frac{d+1}{2^{d-1}\G{\frac{2}{d}+1}} \; c_d^\frac{2}{d} & \text{for odd }d \;.\\ 
 \end{cases}
\end{align}

\subsection{A general calculation for \texorpdfstring{$\frac{\alpha_d}{\beta_d}$}{alphad over betad }}

In \cite{Dowker:2013vba} after observing the pattern of $\alpha_d$ for $d=2,\dots,7$ a closed form expression in all dimensions was conjectured
\begin{equation}
\alpha_d= 
\begin{cases}
\frac{- 2 c_d^{\twod}}{\G{\frac{2}{d}+1}} & \text{if }d\text{ is even} \\
\frac{- c_d^{\twod}}{\G{\frac{2}{d}+1}} & \text{if }d\text{ is odd \;.} 
\end{cases}
\end{equation}
In this section we will calculate the ratio $\frac{\alpha_d}{\beta_d}$, which together with our closed form expressions for $\beta_d$ will allow us to prove above pattern.

\begin{align}
\frac{\alpha_d }{\beta_d}&= - \lim_{l \rightarrow 0} 
 \frac{ S_{d-2}}{l^d}  \Od{d} \int_{-L}^0 \mathrm{d}u \int_u^0  \mathrm{d}v 
\left(\frac{v-u}{\sqrt{2}}\right)^{d-2} 
e^{-l^{-d} V_{0\,d}(u,v)} \;.
\end{align}
We can calculate this in the same manner as $\beta_d$. Here in the main text we will present an outline of the calculations, while more detail is available in appendix \ref{app:alpha}.

We start with the even dimensional case,
\begin{align}
&\frac{\Od{\deve}}{l^d} \intl{-L}{0} \mathrm{d}u \intl{u}{0}  \mathrm{d}v 
\left(\frac{v-u}{\sqrt{2}}\right)^{d-2} 
e^{-l^{-d} c_d (u v)^{\dtwo}} \\
&=\suml{k=0}{d-2} \binom{d-2}{k} \frac{(-1)^k}{2^{\dtwo-1}} \frac{L^{d}}{l^{d+2}\;d (k+1)} \times \nonumber \\
& \bigspace \mFm{\dtwo +1}{\twod(k+1),\twod +1,  \dots, 2- \twod, 2+\frac{2}{d}}{\twod(k+1)+1,\twod , \dots, 1-\twod , 1+\frac{2}{d}}{- c_d \left( \frac{L}{l} \right)^d} \;. \label{eq:sumkevenab}
\end{align}
We introduce $z$ and compare the leading powers of $z$ in 
\begin{align}
z \mFm{\dtwo +1}{\twod(k+1),\twod +1,  \dots, 2- \twod, 2+\frac{2}{d}}{\twod(k+1)+1,\twod ,  \dots ,1-\twod , 1+\frac{2}{d}}{- z}
\end{align}
for the large $z$ limit.
For $k<\dtwo-1$ the first argument of the hypergeometric can be simplified against one of the later arguments. This leads to a function that vanishes in the large $z$ limit according to equation \eqref{eq:limit0}. For $k\geq \dtwo-1$ the function is of the form \eqref{eq:limit}, with a leading order term $ z^{-\twod (k+1)+1}$. This goes to zero for $z \to \infty$ if $k> \dtwo-1$, so the only finite contribution comes from the term $k=\dtwo-1$. Inserting this in equation \eqref{eq:sumkevenab} we find:
\begin{align}
&\binom{d-2}{\dtwo-1} \frac{(-1)^{\dtwo-1}}{2^{\dtwo-2}} \frac{1}{c_d \; d^2}  \frac{\G{2}}{1+\dtwo} \prod\limits_{i=1}^{\dtwo -1} \frac{ i-\dtwo}{i}
= \frac{\binom{d-2}{\dtwo-1}}{2^{\dtwo-3}d^2 (2+d) \; c_d}  
\end{align}
Using this and the definition of $c_d$ (equation \eqref{eq:defc}) in equation \eqref{eq:defalphabeta} leads to
\begin{align}
\frac{\alpha_\deve}{\beta_\deve}&= 
  \frac{S_{d-2} \; \binom{d-2}{\dtwo-1}}{2^{\dtwo-3}d^2 (2+d) \; c_d} 
= - \frac{\G{d}}{\G{\frac{d}{2}+2} \G{\frac{d}{2}}} \label{eq:solalphabetaeve}.
\end{align}
We can combine this with $\beta_\deve$ from equation \eqref{eq:solbetaeven} and find
\begin{align}
\alpha_\deve		&= -\frac{2}{\G{\frac{2}{d}+1}} c_d^{\twod}
\end{align}
as conjectured.

In odd dimensions we find
\begin{align}
&\frac{\Od{\dodd}}{l^d} \intl{-L}{0} \mathrm{d}u \intl{u}{0}  \mathrm{d}v 
\left(\frac{v-u}{\sqrt{2}}\right)^{d-2} 
e^{-l^{-d} c_d (u v)^{\dtwo}} \\
&=\suml{k=0}{d-2} \binom{d-2}{k} \frac{(-1)^k}{2^{\dtwo-1}} \frac{L^{d}}{l^d \; d (k+1)} \times \nonumber \\
& \bigspace  \mFm{\frac{d+1}{2}+2}{1,\twod(k+1),\twod +1,  \dots , 2+\frac{1}{d}}{2,\twod(k+1)+1,\twod , \dots , 1+\frac{1}{d}}{- c_d \left( \frac{L}{l} \right)^d} \;. \label{eq:sumkodddab}
\end{align}
We then introduce $z$ and use \eqref{eq:expident} to expand the hypergeometric function,
\begin{align}
&\lim_{z \to \infty} z \mFm{\frac{d+1}{2}+2}{1,\twod(k+1),\twod +1,  \dots , 2+\frac{1}{d}}{2,\twod(k+1)+1,\twod ,  \dots , 1+\frac{1}{d}}{-z} \\
&=\lim_{z \to \infty} \frac{z}{1-\twod(k+1)}  \bigg\{ \mFm{\frac{d+1}{2}+1}{\twod(k+1),\twod +1, \dots , 2+\frac{1}{d}}{\twod(k+1)+1,\twod ,  \dots , 1+\frac{1}{d}}{- z} \nonumber \\
&\bigspace \ \ \ \ \ \   - \twod (k+1) \mFm{\frac{d+1}{2}+2}{1,\twod +1,  \dots , 2+\frac{1}{d}}{2,\twod ,  \dots , 1+\frac{1}{d}}{- z} \bigg\} \;.
\end{align}
As for $\beta_\dodd$ above the first term does not contribute in the $z \to \infty$ limit.
All terms for $k<\dtwo -1$ can be simplified into a form that vanishes in the large $z$ limit according to equation \eqref{eq:limit0}, while all terms with $k>\dtwo-1$ have a leading order $z^{-\twod(k+1)}$ which falls off faster than $z$ and vanish by equation \eqref{eq:limit}.

We then only need to consider the second term, for which we can take the limit using \eqref{eq:limit}.
Inserting the result of this in equation \eqref{eq:sumkodddab} we can then solve the summation over $k$
\begin{align}
&\suml{k=0}{d-2} \binom{d-2}{k} \frac{(-1)^{k+\frac{d+1}{2}}}{d \; 2^{\dtwo}} \frac{1}{ c_d } \frac{1}{\dtwo -k-1} \frac{  \G{\dtwo}}{\sqrt{\pi} \; \G{\frac{d+3}{2}}}\\
&=\frac{2^{\dtwo}}{d(d-1)(d+1)} \frac{1}{c_d} \;.
\end{align}
This and the definition of $c_d$ in equation \eqref{eq:defc} then lead to
\begin{align}
\frac{\alpha_\dodd }{\beta_\dodd}&= -  S_{d-2} \frac{2^{\dtwo}}{d(d-1)(d+1)} \frac{1}{c_d}
\; =\;  - \frac{2^{d-1}}{d+1} \label{eq:solalphabetaodd}.
\end{align}
We can combine this with $\beta_\dodd$ we find
\begin{equation}
\alpha_\dodd 
= - \frac{c_d^\twod}{\G{\frac{2}{d}+1}} 
\end{equation}
in concordance with the conjecture.

In this section we solved eqn \eqref{eq:defbeta} and \eqref{eq:defalphabeta} to find a closed form expression for $\alpha_d,\beta_d$. The closed form for $\alpha_d$ had been conjectured in \cite{Dowker:2013vba}, but had not been proven before. This closed form is useful for simulations and can be used in further analytic work on causal sets.

\section{\label{sec:c}Finding the \texorpdfstring{$C_i^{(d)}$}{Ci}} 

Knowing a closed form expression for the constants $\alpha_d$ and $\beta_d$ we are one step away from being able to immediately calculate the operator $B^{(d)}$ in any dimension.
The step that is missing is knowing the prefactors $C_i^{(d)}$.
These prefactors are defined as 
\begin{equation}
\Od{d} e^{-l^{-d} V} = \sum_{i =1}^{n_d} C^{(d)}_i \frac{(l^{-d} V)^{i-1}}{(i-1)!} \exp(-l^{-d} V)\;.
\end{equation}
Our task is to find a closed form expression for this. Again we will present a short outline in the main text and go into more detail in appendix \ref{app:Ci}.
In even dimensions we can calculate 
\begin{align}
\Od{\deve} e^{-l^{-d}V}		&= \mFm{\dtwo  +1}{\twod +1, \frac{4}{d}+1,\dots, \twod +2}{\twod , \frac{4}{d},\dots, \twod +1}{-l^{-d}V} \; ,
\end{align}
while in odd dimensions
\begin{align}
\Od{\dodd} e^{-l^{-d}V}		&= \mFm{\frac{d+1}{2}}{\twod +1, \frac{4}{d}+1,\dots, \frac{1}{d}+2}{\twod , \frac{4}{d},\dots, \frac{1}{d}+1}{-l^{-d}V} \; .
\end{align}
Comparing this to equations \eqref{eq:sumeve}, \eqref{eq:oddintermediate}, \eqref{eq:sumkevenab} and \eqref{eq:sumkodddab} above can see that the last $\dtwo+1$ resp. $\frac{d+1}{2}$ terms in these hypergeometric functions are generated by the operator\footnote{The last of these terms in equation \eqref{eq:sumeve} has been simplified away, however this does not change the argument.}. 
In the last section we saw how for odd and even dimensions these terms suppressed contributions that would lead to infinite contributions in the $l \to 0$ limit. The situation is always of the shape
\begin{equation}
z^a \mFm{p+1}{b_k,b_1+1, \dots, b_p+1}{b_k+1,b_1, \dots , b_p}{-z}
\end{equation}
where $b_i$ is a rising sequence and for all $b_k <a$, which would give rise to terms $z^{a-b_1}$ that blow up as $z \to \infty$, the operator generates a new argument that cancels it out.
So the operator generates exactly the terms necessary for a finite result, creating the minimal number of layers $n_d$.


We can then define a generating function for the $C_i^{(d)}$, introducing $z$ 
\begin{align}\label{eq:defGC}
G_\deve &=e^{z} \mFm{\dtwo  +1}{\twod +1, \frac{4}{d}+1,\dots, \twod +2}{\twod , \frac{4}{d},\dots, \twod +1}{-z} & & \text{for even $d$}\\
G_{\dodd}&= e^{z}\mFm{\frac{d+1}{2}}{\twod +1, \frac{4}{d}+1,\dots, \frac{1}{d}+2}{\twod , \frac{4}{d},\dots, \frac{1}{d}+1}{-z} & & \text{for odd $d$}
\end{align}
From these we can generate the $C_i^{(d)}$ as
\begin{equation}
C_i^{(d)}=\left( \pd{}{z} \right)^{i-1} G_d \; \bigg|_{z=0}
\end{equation}
We calculate this in appendix \ref{app:gen} and find:
\begin{align}
C_i^{(\deve)}=\sum_{k=0}^{i-1} \binom{i-1}{k} (-1)^k \, \frac{\G{\dtwo(k+1)+2}}{\G{\dtwo+2}{\G{1+\frac{d k}{2}}}}\\
C_i^{(\dodd)}=\sum_{k=0}^{i-1} \binom{i-1}{k} (-1)^k \, \frac{\G{\dtwo(k+1)+\frac{3}{2}}}{\G{\frac{d+3}{2}}{\G{1+\frac{d k}{2}}}} \;.
\end{align}
These are finite sums that can be evaluated for a given $i$, without having to specify $d$. They can be implemented in computer programs to obtain the action for any possible dimension. 
The computing time for solving the sums should be negligible compared to typical time scales in causal set simulations, therefore these expressions could be used in dimension independent Monte Carlo simulations. Yet this expression remains aesthetically unsatisfying. The hope was to be able to solve the sum over $k$ to a closed form expression for general $i$, to obtain further insights into the structure of causal set space-times.

\section{\label{sec:r}Finding the prefactor of \texorpdfstring{$R$}{R} in all dimensions}
On flat space-times $B^{(d)}$ approximates the d'Alembertian. As mentioned in the introduction, in a curved space-time corrections proportional to the curvature scalar $R$ arise\cite{Benincasa:2010ac},
\begin{align}\label{eq:boxricci}
\lim_{l \rightarrow 0} 
{\bar{B}}^{(d)}\phi(x) &=\Box^{(d)} \phi(x) + a_d R(x)\phi(x)\,,
\end{align}
which can be used to define a causal set action.
In principle the factor $a_d$ could be dimension dependent, but in \cite{Dowker:2013vba} it was shown to be $-\frac{1}{2}$ for $d=2, \dots, 7$. This led to the conjecture that $a_d=-\frac{1}{2}$ for all dimensions.

We will now prove this conjecture.
In \cite{Dowker:2013vba} equation \eqref{eqn:curved} was expanded, to first order in the curvature, in Riemann normal coordinates
\begin{align}
& \beta l^{-(d+2)} S_{d-2} \Od{d}\int_{\mathcal{N}} \!\! \mathrm{d} t\, \mathrm{d}r \, r^{d-2} 
  \nonumber \\
&\bigspace \times \bigg[ \bigg( R_{00} t^2 + \frac{1}{d-1}(R_{00}+ R) r^2\bigg)
             \bigg(-\frac{1}{6} - \frac{1}{24(d+1)}l\frac{\partial}{\partial l}\bigg) e^{-l^{-d}V_{0\,d}}
	\nonumber	\\ 				
             &\bigspaced  + R(t^2 - r^2)\frac{1}{24(d+1)(d+2)} l  \frac{\partial}{\partial l} e^{-l^{-d}V_{0\,d}} \bigg]\,.
\end{align} 
This defines two contributions. The terms proportional to $R$ define $a_d$, while the terms proportional to $R_{00}$ should, due to symmetry, sum up to zero.
\begin{align}
I^{(d)}_R &:= \lim_{l\to 0} \beta l^{-(d+2)} S_{d-2} \nonumber \\
& \ \ \ \ \ \bigg\{ \bigg(-\frac{1}{6} - \frac{1}{24(d+1)}l\frac{\partial}{\partial l}\bigg)   \frac{\Od{d}}{(d-1)}\int_{-L}^0 {\mathrm{d}}u  \!\! \int_{u}^0 \mathrm{d} v \, 
\bigg(\frac{v-u}{\sqrt{2}}\bigg)^{d}   e^{-l^{-d}V_{0\,d}}  \nonumber \\
&\ \ \ \ \ \ \ \ \  +\frac{1}{12(d+1)(d+2)} \; l \frac{\partial}{\partial l}  \Od{d}  \int_{-L}^0 {\mathrm{d}}u  \!\! \int_{u}^0 \mathrm{d} v \, 
\bigg(\frac{v-u}{\sqrt{2}}\bigg)^{d-2} uv\;   e^{-l^{-d}V_{0\,d}} \bigg\} \label{eq:IRdef} \\
I^{(d)}_{00}&: = \lim_{l\to 0} l^{-(d+2)}\beta S_{d-2} \nonumber \\
&\bigspace \bigg\{  \left(-\frac{1}{6} - \frac{1}{24(d+1)} l \pd{}{l} \right)  \frac{\Od{d}}{(d-1)}\intl{-L}{0} \md u \intl{u}{0} \md v \left( \frac{v-u}{\sqrt{2}} \right)^d e^{-l^{-d} V_{0\,d}}  \nonumber \\
& \bigspaced +\left(-\frac{1}{6} - \frac{1}{24(d+1)} l \pd{}{l} \right) \Od{d} \nonumber \\
& \bigspaced \intl{-L}{0} \md u \intl{u}{0} \md v \left( \frac{v-u}{\sqrt{2}} \right)^{d-2} \left( \frac{v+u}{\sqrt{2}} \right)^2 e^{-l^{-d} V_{0\,d}}\bigg\} \,. \label{eq:I0def}
\end{align}
The first term in both of these integrals is the same while the second is different. We will thus have to calculate three different integrals for even and odd dimensions.
The techniques to solve the integrals are the same that have been used throughout this article, so we will only sketch the results here. A more detailed discussion can be found in appendix \ref{app:R}.

We will first calculate all three integrals in even dimensions. 
\begin{align}
&\lim_{l\to 0} l^{-(d+2)}\beta_\deve S_{d-2} \nonumber \\
&\bigspace  \left(-\frac{1}{6} - \frac{1}{24(d+1)} l \pd{}{l} \right)  \frac{\Od{\deve}}{d-1}\intl{-L}{0} \md u \intl{u}{0} \md v \left( \frac{v-u}{\sqrt{2}} \right)^d e^{-l^{-d} V_{0\,d}} \label{eq:eveminusdef}
\end{align}
In the second term applying the derivative $l \pd{}{l}$ changes the arguments of the hypergeometric function
\begin{equation} \label{eq:derivative}
\pd{}{z} \pFq{p}{q}{a_1, \dots ,a_p}{b_1, \dots ,b_q}{z} = \frac{ \prod_{j=1}^{p} a_j}{\prod_{j=1}^q b_q} \pFq{p}{q}{a_1+1, \dots ,a_p+1}{b_1+1, \dots ,b_q+1}{z} \;.
\end{equation}
Using this, as shown in appendix \ref{app:Rodd}, we find
\begin{align}
&=-\frac{1}{3} 		\beta_\deve S_{d-2} \frac{2^{\dtwo-3}\G{\frac{d-1}{2}} \G{\twod+1} }{(d+2)\sqrt{\pi} \G{1+\dtwo}c_d^{1+\twod}} \frac{ 5d+6 }{ d (d+1)} \;.
\end{align}
The other terms  we need to calculate are
\begin{align}
&\lim_{l\to 0} l^{-(d+2)}\beta_\deve S_{d-2} \times \nonumber\\
& \ \ \ \ \ \ \ \   \left(-\frac{1}{6} - \frac{1}{24(d+1)} l \pd{}{l} \right) \Od{\deve}\intl{-L}{0} \md u \intl{u}{0} \md v \left( \frac{v-u}{\sqrt{2}} \right)^{d-2} \left( \frac{v+u}{\sqrt{2}} \right)^2 e^{-l^{-d} V_{0\,d}} \label{eq:eveplusdef}\\
&=\frac{1}{3} \beta_\deve S_{d-2} \frac{2^{\dtwo-3}\G{\frac{d-1}{2}} \G{\twod+1} }{(d+2)\sqrt{\pi} \G{1+\dtwo}c_d^{1+\twod}} \frac{ 5d+6 }{ d (d+1)}\;.
\intertext{and}
&\lim_{l\to 0} \beta_\deve l^{-(d+2)} S_{d-2} \times \nonumber \\
&\bigspace \frac{1}{12(d+1)(d+2)} \; l \frac{\partial}{\partial l}  \Od{\deve}  \int_{-L}^0 {\mathrm{d}}u  \!\! \int_{u}^0 \mathrm{d} v \, 
\bigg(\frac{v-u}{\sqrt{2}}\bigg)^{d-2} uv\;   e^{-l^{-d}V_{0\,d}} \\
 &=-  \frac{ 2^{\dtwo-3}}{3} \frac{d \beta_\deve  S_{d-2}}{(d+1)^2(d+2)^2 \; c_d^{1+\twod}} \frac{\G{3+\twod} \G{\frac{d-1}{2}}}{\G{\dtwo} \sqrt{\pi}}\;.
\end{align}
We can then calculate $I^{(d)}_{00}$ and $I^{(d)}_{R}$ by inserting the constant term
\begin{equation}
\frac{\beta_\deve S_{d-2}}{ c_d^{1+\twod}}= \frac{2^{\dtwo} \; d \; \G{\frac{d}{2}+2} \G{\frac{d}{2}+1}}{\G{\frac{2}{d}} \G{d-1}}
\end{equation}
and the three integrals into equation \eqref{eq:I0def} and \eqref{eq:IRdef}.
\begin{align}
I^{(d)}_{00}&=-\frac{\beta_\deve S_{d-2}}{3} 		 \frac{2^{\dtwo-4}\G{\frac{d-1}{2}} \G{\twod+1} }{\sqrt{\pi} \G{2+\dtwo}c_d^{1+\twod}}\left( \frac{ 5d+6 }{ d (d+1)}-  \frac{ 5d+6 }{ d (d+1)} \right)\\
&=0\\
I^{(d)}_R&=- 	 \frac{ 2^{d-2}}{3 }	\frac{d \; \G{\frac{d}{2}+2}}{ \G{d-1}} \frac{\G{\frac{d-1}{2}}}{(d+1)(d+2)\sqrt{\pi} } \left( \frac{ 5d+6}{d^2 }  +  \frac{1}{d}  \right)\\
&=-\frac{1}{2}
\end{align}
Which proves the conjecture for even dimensions.

The same integrals need to be solved in odd dimensions:
\begin{align}
&\lim_{l\to 0} l^{-(d+2)}\beta_\dodd S_{d-2} \nonumber \\
& \bigspace \left(-\frac{1}{6} - \frac{1}{24(d+1)} l \pd{}{l} \right)  \frac{\Od{d}}{d-1}\intl{-L}{0} \md u \intl{u}{0} \md v \left( \frac{v-u}{\sqrt{2}} \right)^d e^{-l^{-d} V_{0\,d}} \label{eq:oddminusdef}\\
&=-\frac{1}{3} \beta_\dodd S_{d-2} \frac{2^\dtwo \G{\twod+1}}{(d+1)(d-1) c_d^{1+\twod}} \frac{5d+6}{4d(d+1)}\\
&\lim_{l\to 0} l^{-(d+2)}\beta_\dodd S_{d-2} \times \nonumber\\
& \ \ \ \ \ \left(-\frac{1}{6} - \frac{1}{24(d+1)} l \pd{}{l} \right) \Od{d}\intl{-L}{0} \md u \intl{u}{0} \md v \left( \frac{v-u}{\sqrt{2}} \right)^{d-2} \left( \frac{v+u}{\sqrt{2}} \right)^2 e^{-l^{-d} V_{0\,d}} \label{eq:oddplusdef}\\
&=\beta_\dodd S_{d-2} \frac{2^{\dtwo} \G{1+\twod}}{3(d+1)(d-1)} \frac{5d+6}{4 d(d+1)}\\
\intertext{and}
&\lim_{l\to 0} \beta_\dodd l^{-(d+2)} S_{d-2} \times \nonumber \\
&\bigspace \frac{1}{12(d+1)(d+2)} \; l \frac{\partial}{\partial l}  \Od{d}  \int_{-L}^0 {\mathrm{d}}u  \!\! \int_{u}^0 \mathrm{d} v \, 
\bigg(\frac{v-u}{\sqrt{2}}\bigg)^{d-2} uv\;   e^{-l^{-d}V_{0\,d}} \\
&= -\frac{1}{3}\beta_\dodd  S_{d-2} \frac{2^{\dtwo-2} \G{1+\twod}}{(d-1)(d+1)^2 c_d^{1+\twod}}\;.
\end{align}
To calculate $I^{(d)}_{00}$ and $I^{(d)}_{R}$ we insert
\begin{equation}
\frac{\beta_\dodd S_{d-2}}{c_d^{1+\twod}} =  \frac{d(d-1)(d+1)}{2^{\dtwo}\G{\frac{2}{d}+1}} 
\end{equation}
and the integrals into equation \eqref{eq:I0def} and \eqref{eq:IRdef}.
\begin{align}
I^{(d)}_{00}&=  \frac{1}{3} \beta_\dodd S_{d-2} \frac{2^\dtwo \G{\twod+1}}{(d+1)(d-1) c_d^{1+\twod}}  \left(-\frac{5d+6}{4d(d+1)}+ \frac{5d+6}{4 d(d+1)} \right)\\
 &=0\\
\intertext{and}
I^{(d)}_R&= -\frac{1}{3} d \left( \frac{5d+6}{4d (d+1)} + \frac{1}{4(d+1) } \right)\\
&=-\frac{1}{2} \;. 
\end{align}
Which proves the conjecture in odd dimensions. Thus it is  proven that $a_d=-\frac{1}{2}$ in all dimensions.

\section{Conclusion}
In this note we provided closed form expressions for $\alpha_d$ and $\beta_d$ for all dimensions and proved that the coefficient for the scalar curvature $a_d$ will always be $-\frac{1}{2}$.
This completes the work presented in \cite{Dowker:2013vba}. Calculating the $\alpha_d$ and $\beta_d$ also allow some insight into how the operator $\Od{d}$, and thus the layering structure, regularizes the integral over all space. Before applying the operator the integral could be written as a $_2 F_2$ hypergeometric function of the form
\begin{equation}
z^{1+\twod} \mFm{2}{1+\twod, \twod (k+1)}{2+\twod, \twod (k+1)+1}{-z}
\end{equation}
this diverges for the values $k<\dtwo$. The operator $\Od{d}$ creates new arguments (c.f. \eqref{eq:oddop} and \eqref{eq:evenopp}) which cancel these terms out.
We can then say that $\Od{d}$ is the minimal viable operator in the sense that is the operator that generates the minimal number of terms necessary to obtain a finite result. 
One can then presume that the cancellation between the layers that leads to a finite discrete result is the same cancellation that in the continuum approximation leads to a hypergeometric function with a finite infinite density limit.

We also give a closed form expression for the $C_i^{(d)}$ that can easily be evaluated for any given value for $i$.
Our hope is that this work will be of use for further studies of the causal set action in different dimensions.

\section*{Acknowledgments}
I would like to thank Fay Dowker for suggesting this project, Lawrence Phillips, whose M.Sc. thesis gave the starting point for this work, for much correspondence.

The author acknowledges support from the ERC-Advance grant 291092, “Exploring the Quantum Universe” (EQU) and of FNU, the Free Danish Research Council, from the grant “quantum gravity and the role of black holes”.

This research was supported in part by Perimeter Institute for Theoretical Physics. Research at Perimeter Institute is supported by the Government of Canada through Industry
Canada and by the Province of Ontario through the Ministry of Economic Development \& Innovation.
\bibliographystyle{customstyle}
\bibliography{factors_new}
\pagebreak
\begin{appendix}
This article comes with a rather long list of appendices which contain technical details of the calculations presented in the main text. In appendix \ref{app:total} the calculations for $\alpha_d$ and $\beta_d$ are shown in detail, appendix \ref{app:limitident} contains proofs for the hypergeometric function identities that are used to obtain the limit $l \to 0$, in appendix \ref{app:CiAll} some of the technical detail for calculating the constants $C_i^{(d)}$ is presented and lastly appendix \ref{app:R} contains the calculations to prove that $a_d=-\frac{1}{2}$.
\section{\label{app:total}Calculating \texorpdfstring{$\beta_d$ and $\alpha_d$}{betad and alphad}}
In this appendix the more technical details of calculating the constants $\alpha_d$ and $\beta_d$ are presented.
\subsection{\label{app:beta}A general calculation for \texorpdfstring{$\beta_d$}{betad}}

The first step is to expand the exponential and the power $ \left( \frac{v-u}{\sqrt{2}} \right)^{d}$ as power series in $u,v$.
\begin{align}
&\intl{-L}{0} \md u \intl{u}{0} \md v  \left( \frac{v-u}{\sqrt{2}} \right)^{d} e^{-l^{-d}  c_d (uv)^{\dtwo  }} \\
&=\suml{k=0}{d}\binom{d}{k} \frac{(-1)^k}{2^{\dtwo  }} \suml{n=0}{\infty} \frac{\left(-l^{-d}c_d \right)^n}{n!} \intl{0}{L} \md u \intl{0}{u} \md v u^{\dtwo  (n+1)-k}v^{\dtwo  n+k} \\
&=\suml{k=0}{d}\binom{d}{k} \frac{(-1)^k}{2^{\dtwo  }} \suml{n=0}{\infty} \frac{\left(-l^{-d}c_d \right)^n}{n!} \frac{L^{d(n+1)+2}}{(\dtwo  n+k+1)(d(n+1)+2)} \tag{\ref{app:eq:intbeta1}}
\end{align}
Next we apply the operator $\Od{d}$. At this point we will split the discussion into two parts, since the operator differs between odd and even dimensions.

\subsubsection{In even dimensions}
First we apply the operator $\Od{\deve}$ for even dimensions to $l^{-dn}$
\begin{align}
\Od{\deve}l^{-d n}= \frac{1}{2^{\frac{d}{2}+1} (\frac{d}{2}+1)!} \prod\limits_{i=1}^{\frac{d}{2}+1} (2 i -l \pd{}{l}) l^{-dn} = l^{-dn} \prod\limits_{i=1}^{\frac{d}{2}+1} \frac{\Poch{\twod i+1}{n}}{\Poch{\twod  i}{n}} \;. \label{eq:evenopp}
\end{align}
We can then insert this in equation \eqref{eq:intbeta1},
\begin{align}
&\frac{\Od{\deve}}{l^{d+2}} \intl{-L}{0} \md u \intl{u}{0} \md v  \left( \frac{v-u}{\sqrt{2}} \right)^{d} e^{-l^{-d} c_d (uv)^{\dtwo  }}  \\
=&\suml{k=0}{d}\binom{d}{k} \frac{(-1)^k}{2^{\dtwo}} \frac{ L^{2+d}}{l^{d+2} (k+1)(d+2)} \times \nonumber \\
& \bigspace  \suml{n=0}{\infty} \frac{\left(-(\frac{L}{l})^{d} c_d \right)^n}{n!} \frac{\Poch{\twod (k+1)}{n}\Poch{\twod +1}{n}}{\Poch{\twod (k+1)+1}{n}\Poch{\twod +2}{n}} \prod\limits_{i=1}^{\frac{d}{2}+1} \frac{\Poch{\twod i+1}{n}}{\Poch{\twod  i}{n}}\\
=&\suml{k=0}{d}\binom{d}{k} \frac{(-1)^k}{2^{\dtwo}} \frac{L^{2+d}}{l^{d+2} (k+1)(d+2)} \nonumber \\
&\bigspace  \mFm{\dtwo+3}{ 1+\twod,\twod(k+1),\twod +1,\dots, \twod +2 }{ 2+\twod,\twod(k+1)+1,\twod , \dots, \twod +1 }{-c_d \left( \frac{L}{d}\right)^d} \;, \label{eq:eveFmf}
\end{align}
and introduce  $z$:
\begin{align}
=&\suml{k=0}{d}\binom{d}{k} \frac{(-1)^k }{2^{\dtwo} (k+1)(d+2) c^{1+\twod}} z^{1+\twod}   \mFm{\dtwo+1}{\twod(k+1),\twod +1,\dots, 2 }{\twod(k+1)+1,\twod , \dots, 1 }{-z} \;.
\end{align}
To take the limit $z\to \infty$ we need two hypergeometric function identities which are proven in appendix \ref{app:limitident}.
\begin{align}
&\lim_{z\rightarrow \infty} e^{z} \mFm{q}{a_{1},\dots, a_{q}}{a_{1}-1,\dots, a_{q}-1}{-z}  = \frac{(-z)^{q}}{\prod\limits_{j=1}^{q} (a_{j}-1)} +O\left( (-z)^{q-1}\right)  \tag{\ref{eq:limit0}}\\
&	\lim_{z\rightarrow \infty} z^{a_{0}} \mFm{q+1}{a_{0},a_{1}, \dots,a_{q}}{a_{0}+1,a_{1}-1, \dots,a_{q}-1}{-z} = \G{a_{0}+1}  \prod\limits_{j=1}^{q}\frac{a_{j}- a_{0} -1}{a_{j}-1} \tag{\ref{eq:limit}}
\end{align}
In even dimensions when taking the limit $z\to \infty$ we split the terms into three categories depending on $k$.
For those terms where $k < \dtwo$ the first term in the hypergeometric function can be simplified against one of the later terms, leaving us with a function of the form \eqref{eq:limit0}. Thus the terms for $k<\dtwo$ do not contribute. If $k\geq \dtwo$ no simplification occurs and the hypergeometric function is of the form \eqref{eq:limit} with leading order term $z^{-\twod(k+1)}$. For $k>\dtwo$ this falls of faster than $z^{1+\twod}$, therefore these terms do not contribute. The only finite contribution in the limit is then from the term $k=\dtwo$.
\begin{align}
\lim_{z\to \infty} &\binom{d}{\dtwo} \frac{(-1)^\dtwo}{2^{\dtwo} (\dtwo+1)(d+2) c^{1+\twod}} z^{1+\twod}  \nonumber \\
&\bigspace \mFm{\dtwo+1}{\twod+1, \twod +1,\dots, 2 }{\twod+2,\twod , \dots, 1 }{-z}\\
=&\binom{d}{\dtwo} \frac{(-1)^\dtwo }{2^{\dtwo} (\dtwo+1)(d+2) c^{1+\twod}} \G{2+\twod} \prod\limits_{i=1}^\dtwo  \frac{ i-1-\dtwo}{i} \\ \label{eq:evelimitfin}
=&\binom{d}{\dtwo} \frac{1}{2^{\dtwo-1}d (d+2) c^{1+\twod}} \G{\frac{2+d}{d}} 
\end{align}
We can insert this into equation \eqref{eq:defbeta} and find
\begin{align}
	\beta_\deve &=\frac{2 \; \G{\frac{d}{2}+2} \G{\frac{d}{2}+1}}{\G{\frac{2}{d}} \G{d}} \; c_d^{\twod} 
\end{align}

\subsubsection{\label{app:subsubbodd}In odd dimensions}
Equation \eqref{eq:intbeta1} depends on $l$ as $l^{-dn}$. Applying the operator $\Od{\dodd}$ then gives
\begin{align} \label{eq:oddop}
\Od{\dodd}l^{-d n}= \frac{1}{2^{\frac{d+1}{2}} (\frac{d+1}{2})!} \prod\limits_{i=1}^{\frac{d+1}{2}} (2 i -l \pd{}{l}) l^{-dn} = l^{-dn} \prod\limits_{i=1}^\frac{d+1}{2} \frac{\Poch{\twod i+1}{n}}{\Poch{\twod  i}{n}} \;.
\end{align}
We  rewrite the denominators containing $n$ as Pochhammer symbols and find,
\begin{align}
&\frac{1}{l^{d+2}}\Od{\dodd} \intl{-L}{0} \md u \intl{u}{0} \md v  \left( \frac{v-u}{\sqrt{2}} \right)^{d} e^{-l^{-d} c_d (uv)^{\dtwo  }}  \\
=&\suml{k=0}{d}\binom{d}{k} \frac{(-1)^k}{2^\dtwo \; l^{d+2}} \frac{L^{2+d}}{(k+1)(d+2)}  \nonumber\\
&\bigspace \suml{n=0}{\infty} \frac{\left(-(\frac{L}{l})^{d} c_d \right)^n}{n!} \frac{\Poch{\twod (k+1)}{n}\Poch{\twod +1}{n}}{\Poch{\twod (k+1)+1}{n}\Poch{\twod +2}{n}} \prod\limits_{i=1}^\frac{d+1}{2} \frac{\Poch{\twod i+1}{n}}{\Poch{\twod  i}{n}}\\
=& \suml{k=0}{d}\binom{d}{k} \frac{(-1)^k}{2^{\dtwo} \; l^{d+2}} \frac{L^{2+d}}{ (k+1)(d+2)} \nonumber \\
& \bigspace \mFm{\frac{d+1}{2}+2}{1+\twod,\twod (k+1), \twod+1, \dots, \frac{1}{d}+2}{2+\twod,\twod (k+1)+1, \twod, \dots, \frac{1}{d}+1}{-c_d (\frac{L}{l})^d}\;. \\
=& \suml{k=0}{d}\binom{d}{k} \frac{(-1)^k}{2^{\dtwo}} \frac{ z^{1+\dtwo} }{ (k+1)(d+2)c_d^{1+\dtwo}} \nonumber \\
& \bigspace \mFm{\frac{d+1}{2}+2}{1+\twod,\twod (k+1), \twod+1, \dots, \frac{1}{d}+2}{2+\twod,\twod (k+1)+1, \twod,  \dots, \frac{1}{d}+1}{-z} \;.\label{eq:oddinserthere}
\end{align}
The limit $l \to 0$ is equivalent to the limit $z \to \infty$. 
Before using equations \eqref{eq:limit0} and \eqref{eq:limit} we will simplify the hypergeometric function once more using \cite{07.31.03.0008.01}
\begin{align}
&\pFq{p}{q}{a,b,a_3,\dots,a_p}{a+1,b+1,b_3,\dots b_q}{z} =  \frac{1}{a-b} \times  \nonumber \\
&\left\{ a \pFq{p-1}{q-1}{b,a_3,\dots,a_p}{b+1,b_3,\dots,b_q}{z} - b \pFq{p-1}{q-1}{a,a_3,\dots,a_p}{a+1,b_3,\dots,b_q}{z} \right\} \label{eq:expident} \;.
\end{align}
This leads to
\begin{align}
&\lim_{z\rightarrow \infty} z^{1+\twod} \mFm{\frac{d+1}{2}+2}{1+\twod,\twod (k+1), \twod+1,  \dots, \frac{1}{d}+2}{2+\twod,\twod (k+1)+1, \twod, \dots, \frac{1}{d}+1}{-z}\\
&=\lim_{z\rightarrow \infty} \frac{z^{1+\twod}}{1-\twod k} \bigg\{
 (1+\twod) \mFm{\frac{d+1}{2}+1}{\twod (k+1), \twod+1,  \dots, \frac{1}{d}+2}{\twod (k+1)+1, \twod,  \dots, \frac{1}{d}+1}{-z}  \nonumber \\
& \bigspace  - \twod (k+1) \mFm{\frac{d+1}{2}+1}{1+\twod , \twod+1,  \dots, \frac{1}{d}+2}{2+\twod , \twod,  \dots, \frac{1}{d}+1}{-z}  \bigg\}
\end{align}
Comparing to equations \eqref{eq:limit0} and \eqref{eq:limit} we can see that the first term does not contribute.  
For $\twod (k+1)< 1+\frac{1}{d}$, all terms where $k< \dtwo$, the term $\twod(k+1)$ can be simplified against one of the later terms, and thus the hypergeometric function goes to zero as $z \to \infty$ \eqref{eq:limit0}. For $k>\dtwo$ ,the leading order behavior can be determined as $z^{-\twod(k+1)}$ using \eqref{eq:limit}. This goes to zero faster than $z^{1+\twod}$, thus also vanishes in the limit.
This leaves us with the second term
\begin{align}
 -\lim_{z\rightarrow \infty} \frac{\twod (k+1)}{1-\twod k} z^{1+\twod}  \mFm{\frac{d+1}{2}+1}{1+\twod , \twod+1, \dots, \frac{1}{d}+2}{2+\twod , \twod,  \dots, \frac{1}{d}+1}{-z}  \;.
\end{align}
We can use equation \eqref{eq:limit} to take the limit $z \to \infty$, leading to
\begin{align}
=&\frac{-(k+1)}{\dtwo-k} \G{2+\twod} \prod\limits_{i=1}^{\frac{d+1}{2}} \frac{i-1-\dtwo}{i} \\
=& \frac{(k+1)}{\dtwo-k} (-1)^{\frac{d+1}{2}+1} \frac{\G{2+\twod} \G{1+\dtwo}}{\G{\frac{d+1}{2}+1} \sqrt{\pi}} \;.
\end{align}
Inserting this into equation \eqref{eq:oddinserthere} above we can now solve the summation over $k$
\begin{align}
=& \suml{k=0}{d}\binom{d}{k} \frac{(-1)^{\frac{d+1}{2}+1+k}}{2^{\dtwo} (\dtwo-k)} \frac{l^{2+d} z^{1+\dtwo} }{ (d+2)c_d^{1+\dtwo}}   \frac{\G{2+\twod} \G{1+\dtwo}}{\G{\frac{d+1}{2}+1} \sqrt{\pi}}  \\
=& \frac{2^{1+\dtwo} \G{\frac{2}{d}+1}}{d(d+1)}\frac{l^{2+d}  }{c_d^{1+\dtwo}} 
\end{align}
This and the definition of $\beta_d$ above \eqref{eq:defbeta}, then allow us to find a closed form expression for $\beta_\dodd$
\begin{align}
\beta_{d_{\text{odd}}}= \frac{d+1}{2^{d-1}\G{\frac{2}{d}+1}} \; c_d^\twod \;. 
\end{align}

\subsection{\label{app:alpha}A general calculation for \texorpdfstring{$\frac{\alpha_d}{\beta_d}$}{alphad over betad }}
In this subsection we will present the calculations to solve equation \eqref{eq:defalphabeta}.
\begin{align}
\frac{\alpha_d }{\beta_d}&= - \lim_{l \rightarrow 0} 
 \frac{ S_{d-2}}{l^d}  \Od{d} \int_{-L}^0 \mathrm{d}u \int_u^0  \mathrm{d}v 
\left(\frac{v-u}{\sqrt{2}}\right)^{d-2} 
e^{-l^{-d} V_{0\,d}(u,v)} \;.
\end{align}
We expand the integrals as a power series in $u$ and $v$:
\begin{align}
& \intl{-L}{0} \mathrm{d}u \intl{u}{0}  \mathrm{d}v 
\left(\frac{v-u}{\sqrt{2}}\right)^{d-2} 
e^{-l^{-d} c_d (u v)^{\dtwo}} \\
&=\suml{k=0}{d-2} \binom{d-2}{k} \frac{(-1)^k}{2^{\dtwo-1}} \suml{n=0}{\infty} \frac{\left(-l^{-d}c_d \right)^n}{n!} \intl{-L}{0} \mathrm{d}u \intl{u}{0}  \mathrm{d}v u^{\dtwo(n+2)-k-2} v^{\dtwo n+k}\\
&=\suml{k=0}{d-2} \binom{d-2}{k} \frac{(-1)^k}{2^{\dtwo-1}} \suml{n=0}{\infty} \frac{\left(-l^{-d}c_d \right)^n}{n!} \frac{1}{(\dtwo n +k+1)d(n+1)} L^{d(n+1)} \;.
\end{align}
Since the terms are proportional to $l^{-dn}$ we can use equation \eqref{eq:oddop} (resp. \eqref{eq:evenopp}) for odd (resp. even) dimensions. 

\subsubsection{In even dimensions}
For the even dimensional case applying $\Od{\deve}$ \eqref{eq:evenopp} leads to
\begin{align}
&\frac{\Od{\deve}}{l^d} \intl{-L}{0} \mathrm{d}u \intl{u}{0}  \mathrm{d}v 
\left(\frac{v-u}{\sqrt{2}}\right)^{d-2} 
e^{-l^{-d} c_d (u v)^{\dtwo}} \\
=&\suml{k=0}{d-2} \binom{d-2}{k} \frac{(-1)^k}{2^{\dtwo-1}} \frac{L^{d}}{l^{d+2} \; d (k+1)} \times \nonumber  \\
& \bigspace \suml{n=0}{\infty} \frac{\left(-c_d \left(\frac{L}{l}\right)^{d}  \right)^d}{n!}  
\frac{ \Poch{1}{n} \Poch{\twod(k+1)}{n}}{ \Poch{2}{n} \Poch{\twod(k+1)+1}{n}} \prod\limits_{i=1}^{\dtwo +1} \frac{\Poch{\twod i+1}{n}}{\Poch{\twod i}{n}}\\
&=\suml{k=0}{d-2} \binom{d-2}{k} \frac{(-1)^k}{2^{\dtwo-1}} \frac{L^{d}}{l^{d+2}\;d (k+1)} \times \nonumber \\
& \bigspace \mFm{\dtwo +1}{\twod(k+1),\twod +1,  \dots, 2- \twod, 2+\frac{2}{d}}{\twod(k+1)+1,\twod , \dots, 1-\twod , 1+\frac{2}{d}}{- c_d \left( \frac{L}{l} \right)^d} \;. 
\end{align}
We introduce $z$ and compare the leading powers for the large $z$ limit.
For $k<\dtwo-1$ the first argument of the hypergeometric can be simplified against one of the later arguments, leading to a function that vanishes in the large $z$ limit according to equation \eqref{eq:limit0}. For $k\geq \dtwo-1$ it is of the form \eqref{eq:limit}, with a leading order term $ z^{-\twod (k+1)+1}$. This goes to zero for $z \to \infty$ if $k> \dtwo-1$, so the only finite contribution comes from the term $k=\dtwo-1$. Inserting this in equation \eqref{eq:sumkevenab} we find:
\begin{align}
&\binom{d-2}{\dtwo-1} \frac{(-1)^{\dtwo-1}}{2^{\dtwo-2}} \frac{1}{c_d \; d^2}  \frac{\G{2}}{1+\dtwo} \prod\limits_{i=1}^{\dtwo -1} \frac{ i-\dtwo}{i}
= \frac{\binom{d-2}{\dtwo-1}}{2^{\dtwo-3}d^2 (2+d) \; c_d}  
\end{align}
Using this and the definition of $c_d$ (equation \eqref{eq:defc}) in equation \eqref{eq:defalphabeta} leads to
\begin{align}
\frac{\alpha_\deve}{\beta_\deve}&= 
  \frac{S_{d-2} \; \binom{d-2}{\dtwo-1}}{2^{\dtwo-3}d^2 (2+d) \; c_d}  
= - \frac{\G{d}}{\G{\frac{d}{2}+2} \G{\frac{d}{2}}} 
\end{align}

\subsubsection{In odd dimensions}
We shall first calculate the odd dimensional case, using equation \eqref{eq:oddop}
\begin{align}
&\frac{\Od{\dodd}}{l^d} \intl{-L}{0} \mathrm{d}u \intl{u}{0}  \mathrm{d}v 
\left(\frac{v-u}{\sqrt{2}}\right)^{d-2} 
e^{-l^{-d} c_d (u v)^{\dtwo}} \\
=&\suml{k=0}{d-2} \binom{d-2}{k} \frac{(-1)^k}{2^{\dtwo-1}} \frac{L^{d}}{ l^{d} \; d (k+1)} \times \nonumber  \\
&\bigspace \suml{n=0}{\infty} \frac{\left(-c_d \left(\frac{L}{l}\right)^{d}  \right)^n}{n!}  
\frac{ \Poch{1}{n} \Poch{\twod(k+1)}{n}}{ \Poch{2}{n} \Poch{\twod(k+1)+1}{n}} \prod\limits_{i=1}^{\frac{d+1}{2}} \frac{\Poch{\twod i+1}{n}}{\Poch{\twod i}{n}}\\
&=\suml{k=0}{d-2} \binom{d-2}{k} \frac{(-1)^k}{2^{\dtwo-1}} \frac{L^{d}}{l^d \; d (k+1)} \times \nonumber \\
& \bigspace  \mFm{\frac{d+1}{2}+2}{1,\twod(k+1),\twod +1,  \dots , 2+\frac{1}{d}}{2,\twod(k+1)+1,\twod , \dots , 1+\frac{1}{d}}{- c_d \left( \frac{L}{l} \right)^d} \;. 
\end{align}
We introduce $z$ and use \eqref{eq:expident} to expand the hypergeometric function.
\begin{align}
&\lim_{z \to \infty} z \mFm{\frac{d+1}{2}+2}{1,\twod(k+1),\twod +1,  \dots , 2+\frac{1}{d}}{2,\twod(k+1)+1,\twod ,  \dots , 1+\frac{1}{d}}{-z} \\
&=\lim_{z \to \infty} \frac{z}{1-\twod(k+1)}  \bigg\{ \mFm{\frac{d+1}{2}+1}{\twod(k+1),\twod +1, \dots , 2+\frac{1}{d}}{\twod(k+1)+1,\twod ,  \dots , 1+\frac{1}{d}}{- z} \nonumber \\
&\bigspace \ \ \ \ \ \   - \twod (k+1) \mFm{\frac{d+1}{2}+2}{1,\twod +1,  \dots , 2+\frac{1}{d}}{2,\twod ,  \dots , 1+\frac{1}{d}}{- z} \bigg\}
\end{align}
As in the case for $\beta_\dodd$ above the first term does not contribute.
All terms for $k<\dtwo -1$ can be simplified into a form that vanishes in the large $z$ limit according to equation \eqref{eq:limit0}, while all terms with $k>\dtwo-1$ have a leading order $z^{-\twod(k+1)}$ which falls off faster than $z$ and vanish by equation \eqref{eq:limit}.
We then only need to consider the second term, which can be expanded using equation \eqref{eq:limit}. 
\begin{align}
&\lim_{z \to \infty} \frac{- z (k+1) }{\dtwo- k-1} \mFm{\frac{d+1}{2}+2}{1,\twod +1,  \dots , 2+\frac{1}{d}}{2,\twod ,  \dots , 1+\frac{1}{d}}{- z} \\
& = \frac{- (k+1)}{\dtwo -k-1} \G{2} \prod\limits_{i=1}^{\frac{d+1}{2}} \frac{ i-\dtwo}{i} 
\;  = \; \frac{(-1)^{\frac{d+1}{2}} (k+1)}{\dtwo -k-1} \frac{\G{\dtwo}}{2 \sqrt{\pi} \G{\frac{d+3}{2}}} \;.
\end{align}
We can reinsert this in equation \eqref{eq:sumkodddab} and find:
\begin{align}
&\suml{k=0}{d-2} \binom{d-2}{k} \frac{(-1)^{k+\frac{d+1}{2}}}{d \; 2^{\dtwo}} \frac{1}{ c_d } \frac{1}{\dtwo -k-1} \frac{  \G{\dtwo}}{\sqrt{\pi} \; \G{\frac{d+3}{2}}}\\
&=\frac{2^{\dtwo}}{d(d-1)(d+1)} \frac{1}{c_d} \;.
\end{align}
This and the definition of $c_d$ in equation \eqref{eq:defc} can be inserted into equation \eqref{eq:defalphabeta} and lead to
\begin{align}
\frac{\alpha_\dodd }{\beta_\dodd}&= -  S_{d-2} \frac{2^{\dtwo}}{d(d-1)(d+1)} \frac{1}{c_d} 
\; =\;  - \frac{2^{d-1}}{d+1} 
\end{align}

\section{\label{app:limitident} Proof of some hypergeometric limits}
We want to prove equation \eqref{eq:limit0}:
\begin{align}
\lim_{z\rightarrow \infty} e^{z} \mFm{q}{a_{1},\dots, a_{q}}{a_{1}-1,\dots, a_{q}-1}{-z}= \frac{(-z)^{q}}{\prod\limits_{j=1}^{q} (a_{j}-1)} +O\left( (-z)^{q-1}\right) 
\end{align}
it is easy to establish that this is true for $q=1,2,3...$ using Mathematica. We can then prove this identity for all $q \in \mathbb{Z}$ by induction.
\begin{align}
	&\lim_{z\rightarrow \infty} e^{z} \mFm{q+1}{a_{1},\dots, a_{q+1}}{a_{1}-1,\dots, a_{q+1}-1}{-z}\\
	&\underset{\text{using the identity \cite{wolfram}}}{=} \lim_{z\rightarrow \infty}\bigg\{ e^{z} \mFm{q}{a_{1},\dots, a_{q}}{a_{1}-1,\dots, a_{q}-1}{-z} +\nonumber \\
	&\bigspace  \frac{\prod\limits_{j=1}^{q}a_{j}\; (-z) e^{z}}{(a_{q+1}-1) \prod\limits_{j=1}^{q}(a_{j}-1)}   \mFm{q}{a_{1}+1,\dots, a_{q}+1}{a_{1},\dots, a_{q}}{-z}  \bigg\} \\
	\intertext{We can now use that identity \eqref{eq:limit0} is true for $q$}
	&= \frac{(-z)^{q}}{\prod\limits_{j=1}^{q} (a_{j}-1)}  +O\left( (-z)^{q-1}\right)  \nonumber \\
	& \bigspace + \frac{\prod\limits_{j=1}^{q}a_{j}\;  }{(a_{q+1}-1) \prod\limits_{j=1}^{q}(a_{j}-1)} \left( \frac{(-z)^{q+1}}{\prod\limits_{j=1}^{q} (a_{j})} +O\left( (-z)^{q}\right) \right)\\
	&= \frac{(-z)^{q+1}}{\prod\limits_{j=1}^{q+1} (a_{j}-1)} +O\left( (-z)^{q}\right) 
\end{align}

The other identity we need to prove is equation \eqref{eq:limit}:
\begin{align}
	&\lim_{z\rightarrow \infty} z^{a_{0}} \mFm{q+1}{a_{0},a_{1}, \dots,a_{q}}{a_{0}+1,a_{1}-1, \dots,a_{q}-1}{-z}
= \G{a_{0}+1}  \prod\limits_{j=1}^{q}\frac{a_{j}- a_{0} -1}{a_{j}-1} \;.
\end{align}
It is easy to establish that this is true for $q=1,2,3...$ using Mathematica. We can then prove this identity for all $q \in \mathbb{Z}$ by induction.
\begin{align}
		&\lim_{z\rightarrow \infty} z^{a_{0}} \mFm{q+2}{a_{0},a_{1}, \dots,a_{q},a_{q+1}}{a_{0}+1,a_{1}-1, \dots,a_{q}-1,a_{q+1}}{-z}\\
		&\underset{\text{using the identity \cite{wolfram}}}{=}  \lim_{z\rightarrow \infty} \bigg\{ z^{a_{0}} \mFm{q+1}{a_{0},a_{1}, \dots,a_{q}}{a_{0}+1,a_{1}-1, \dots,a_{q}-1}{-z} \\
		& \ \ \ \ \ \ \ \ \ \ \ \ \ -\frac{\prod\limits_{j=0}^{q} a_{j} \quad  z^{a_{0}+1} \mFm{q+1}{a_{0}+1,a_{1}+1, \dots,a_{q}+1}{a_{0}+2,a_{1}, \dots,a_{q}}{-z} }{(a_{q+1}-1)(a_{0}+1) \prod\limits_{j=1}^{q}(a_{j}-1)}   \bigg\}\\
		\intertext{We can now use that identity \eqref{eq:limit} is true for $q$}
		&=  \G{a_{0}+1}  \prod\limits_{j=1}^{q}\frac{a_{j}- a_{0} -1}{a_{j}-1} - \frac{\prod\limits_{j=0}^{q} a_{j} \quad \G{a_{0}+2}  \prod\limits_{j=1}^{q}\frac{a_{j}- a_{0} -1}{a_{j}}}{(a_{q+1}-1)(a_{0}+1) \prod\limits_{j=1}^{q}(a_{j}-1)} \\
		&=  \G{a_{0}+1}  \prod\limits_{j=1}^{q}\frac{a_{j}- a_{0} -1}{a_{j}-1} \left( 1- \frac{a_{0}}{a_{q+1}-1}\right)\\
		&= \G{a_{0}+1}  \prod\limits_{j=1}^{q+1}\frac{a_{j}- a_{0} -1}{a_{j}-1} 
\end{align}

\section{\label{app:CiAll}Calculating the \texorpdfstring{$C_i^{(d)}$}{Ci} }
This appendix will show some more detail for the calculations in section \ref{sec:c}.
\subsection{\label{app:Ci}Applying the operator to \texorpdfstring{$e^{-l^{-d}V}$}{Exp(-V/l)} }
In even dimensions we have
\begin{align}
\Od{d} e^{-l^{-d}V}& = \frac{1}{2^{\dtwo  +1} (\dtwo  +1)!} \prod\limits_{i=1}^{\dtwo  +1} \left(-l \pd{}{l}+2 i \right) e^{-l^{-d}V} \\
				&= \frac{1}{2^{\dtwo  +1} (\dtwo  +1)!} \sum\limits_{n=0}^{\infty} \frac{(-V)^n}{n!} \prod\limits_{i=1}^{\dtwo  +1} \left(-l \pd{}{l}+2 i \right) l^{-dn} \\
				&= \frac{d^{\dtwo  +1}}{2^{\dtwo  +1}(\dtwo  +1)!} \sum\limits_{n=0}^{\infty} \frac{(-l^{-d} V)^n}{n!} \prod\limits_{i=1}^{\dtwo  +1} \left( n+\twod  i \right) \\
				&= \sum\limits_{n=0}^{\infty} \frac{(-l^{-d} V)^n}{n!} \prod\limits_{i=1}^{\dtwo  +1} \frac{\Poch{\twod i+1}{n}}{\Poch{\twod i}{n}} \\
				&= \mFm{\dtwo  +1}{\twod +1, \frac{4}{d}+1,\dots, \twod +2}{\twod , \frac{4}{d},\dots, \twod +1}{-l^{-d}V} \; ,
\end{align}
while in odd dimensions
\begin{align}
\Od{d} e^{-l^{-d}V}& = \frac{1}{2^{\frac{d+1}{2}} (\frac{d+1}{2})!} \prod\limits_{i=1}^{\frac{d+1}{2}} \left(-l \pd{}{l}+2 i \right) e^{-l^{-d}V} \\
				&= \frac{1}{2^{\frac{d+1}{2}} (\frac{d+1}{2})!} \sum\limits_{n=0}^{\infty} \frac{(-V)^n}{n!} \prod\limits_{i=1}^{\frac{d+1}{2}} \left(-l \pd{}{l}+2 i \right) l^{-dn} \\
				&= \frac{d^{\frac{d+1}{2}}}{2^{\frac{d+1}{2}}(\frac{d+1}{2})!} \sum\limits_{n=0}^{\infty} \frac{(-l^{-d} V)^n}{n!} \prod\limits_{i=1}^{\frac{d+1}{2}} \left( n+\twod  i \right) \\
				&= \sum\limits_{n=0}^{\infty} \frac{(-l^{-d} V)^n}{n!} \prod\limits_{i=1}^{\frac{d+1}{2}} \frac{\Poch{\twod i+1}{n}}{\Poch{\twod i}{n}} \\
				&= \mFm{\frac{d+1}{2}}{\twod +1, \frac{4}{d}+1,\dots, \frac{1}{d}+2}{\twod , \frac{4}{d},\dots, \frac{1}{d}+1}{-l^{-d}V} \; .
\end{align}
\subsection{\label{app:gen}The generating function}
In section \ref{sec:c} we define the generating function from which to find the $C_i^{(d)}$ in eqn \eqref{eq:defGC}.
Here we calculate an identity to find it's derivatives, starting from the general form  $\mFm{p}{a_1+1, \dots a_p+1}{a_1,\dots, a_p}{-z}$;
\begin{align}
&\left( \pd{}{z}\right)^{n} e^z  \mFm{p}{a_1+1, \dots a_p+1}{a_1,\dots, a_p}{-z}\\
=&\sum\limits_{k=0}^{n} \binom{n}{k} e^z \pd{^k}{z^k}\mFm{p}{a_1+1, \dots a_p+1}{a_1,\dots, a_p}{-z}\\
=&\sum\limits_{k=0}^{n} \binom{n}{k}(-1)^k \, e^z  \,\frac{\prod\limits_{j=1}^p \Poch{a_j+1}{k} }{\prod\limits_{j=1}^p \Poch{a_j}{k} } \mFm{p}{a_1+k+1, \dots a_p+k+1}{a_1,\dots, a_p}{-z}\\
=&\sum\limits_{k=0}^{n} \binom{n}{k}(-1)^k \, e^z   \, \prod\limits_{j=1}^p \frac{a_j+k}{a_j} \mFm{p}{a_1+k+1, \dots a_p+k+1}{a_1,\dots, a_p}{-z} \;.
\end{align}
For $z=0$ this is
\begin{align}
=&\sum\limits_{k=0}^{n} \binom{n}{k} (-1)^k \, \prod\limits_{j=1}^p \frac{a_j+k}{a_j}
\end{align}
Inserting $a_j=\twod j$ and $p= \dtwo+1$ for even $d$ (resp. $p= \frac{d+1}{2}$ for odd $d$) we find
\begin{align}
C_i^{(\deve)}=\sum_{k=0}^{i-1} \binom{i-1}{k} (-1)^k \, \frac{\G{\dtwo(k+1)+2}}{\G{\dtwo+2}{\G{1+\frac{d k}{2}}}}\\
C_i^{(\dodd)}=\sum_{k=0}^{i-1} \binom{i-1}{k} (-1)^k \, \frac{\G{\dtwo(k+1)+\frac{3}{2}}}{\G{\frac{d+3}{2}}{\G{1+\frac{d k}{2}}}} \;.
\end{align}

\section{\label{app:R}Proving the prefactor of \texorpdfstring{$R$}{R} in all dimensions}
In this appendix we calculate the results presented in section \ref{sec:r} of the main text.

\subsection{\label{app:Reven}Even dimensions}
We first solve the integrals in even dimensions.
\begin{align}
&\lim_{l\to 0} l^{-(d+2)}\beta_\deve S_{d-2} \nonumber \\
&\bigspace  \left(-\frac{1}{6} - \frac{1}{24(d+1)} l \pd{}{l} \right)  \frac{\Od{d}}{d-1}\intl{-L}{0} \md u \intl{u}{0} \md v \left( \frac{v-u}{\sqrt{2}} \right)^d e^{-l^{-d} V_{0\,d}}
\end{align}
The integral with $\Od{d}$ acting on it has been calculated  in\eqref{eq:eveFmf} above.
\begin{align}
&=\lim_{l\to 0} l^{-(d+2)}\beta_\deve \frac{S_{d-2}}{d-1} \nonumber \\
& \bigspace \left(-\frac{1}{6} - \frac{1}{24(d+1)} l \pd{}{l} \right)
\suml{k=0}{d}\binom{d}{k} \frac{(-1)^k}{2^{\dtwo}} \frac{L^{2+d}}{l^{d+2} (k+1)(d+2)} \nonumber \\
&\bigspaced \mFm{\dtwo+1}{\twod(k+1),\twod +1,\dots, 2 }{\twod(k+1)+1,\twod , \dots, 1 }{-c_d \left( \frac{L}{d}\right)^d} \;
\end{align}
For the term multiplied by $-\frac{1}{6}$ we can use the limit \eqref{eq:evelimitfin}
\begin{align}
-\frac{1}{3} \binom{d}{\dtwo} \frac{\beta_\deve S_{d-2}}{2^{\dtwo-1}d (d+2)(d-1) c^{1+\twod}} \G{\frac{2+d}{d}}  \\
= 	-\frac{1}{3} 		\beta_\deve S_{d-2} \frac{2^{\dtwo-1} \G{\twod+1} \G{\frac{d-1}{2}}}{d(d+2)\sqrt{\pi} \G{\dtwo+1} c_d^{1+\twod}}  \;.
\end{align}
For the other term we take the derivative
\begin{align}
&\lim_{l\to 0} \frac{-l^{-(d+2)}\beta_\deve S_{d-2}}{24(d+1)(d-1)} l \pd{}{l} 
\suml{k=0}{d}\binom{d}{k} \frac{(-1)^k}{2^{\dtwo}} \frac{L^{2+d}}{l^{d+2} (k+1)(d+2)} \nonumber \\
&\bigspace \mFm{\dtwo+1}{\twod(k+1),\twod +1,\dots, 2 }{\twod(k+1)+1,\twod ,\dots, 1 }{-c_d \left( \frac{L}{d}\right)^d} \; \\
&=\lim_{l\to 0} \frac{- l^{-(d+2)} \beta_\deve S_{d-2}}{24(d+1)(d-1)} \sum_{k=0}^d \binom{d}{k} \frac{(-1)^k \; L^{d+2}}{2^\dtwo (k+1+\dtwo)(d+2)} \prod_{j=1}^{\dtwo} \frac{j+\dtwo}{j} \nonumber \\
&\bigspace \mFm{\dtwo+1}{\twod(k+1)+1,\twod +2,\dots, 3 }{ \twod(k+1)+2,\twod+1 , \dots,2 }{-c_d \left( \frac{L}{d}\right)^d} \;.
\end{align}
Introducing $z$ and taking the limit $z \to \infty$ only the term $k= \dtwo$ contributes, leading to
\begin{align}
-\frac{1}{3} \beta_\deve S_{d-2} \frac{2^{\dtwo-4} \G{\twod+3} \G{\frac{d-1}{2}} d}{(d+1)^2(d+2) \sqrt{\pi} \G{1+\dtwo}c_d^{1+\twod}} \;.
\end{align}
Summing these two we find the solution to \eqref{eq:eveminusdef}
\begin{align}
-\frac{1}{3} 		\beta_\deve S_{d-2} \frac{2^{\dtwo-3}\G{\frac{d-1}{2}} \G{\twod+1} }{(d+2)\sqrt{\pi} \G{1+\dtwo}c_d^{1+\twod}} \frac{ 5d+6 }{ d (d+1)} \;.
\end{align}
The next term  is
\begin{align}
&\lim_{l\to 0} l^{-(d+2)}\beta_\deve S_{d-2} \times \nonumber\\
& \ \ \ \ \ \ \ \   \left(-\frac{1}{6} - \frac{1}{24(d+1)} l \pd{}{l} \right) \Od{d}\intl{-L}{0} \md u \intl{u}{0} \md v \left( \frac{v-u}{\sqrt{2}} \right)^{d-2} \left( \frac{v+u}{\sqrt{2}} \right)^2 e^{-l^{-d} V_{0\,d}} 
\\
&=\lim_{l\to 0} l^{-(d+2)}\beta_\deve S_{d-2} \left(-\frac{1}{6} - \frac{1}{24(d+1)} l \pd{}{l} \right) \nonumber \\
& \bigspace \sum_{x=0}^d \binom{2}{x} \sum_{k=0}^{d-2} \binom{d-2}{k} \frac{(-1)^k}{2^{\dtwo}} \frac{L^{d+2}}{(k+x+1)(d+2)} \nonumber \\
& \bigspace \ \ \ \ \ \ \ \mFm{\dtwo +1}{\twod (k+x+1),\twod+1,\dots,2}{\twod (k+x+1)+1,\twod,\dots,1}{-c_d \left(\frac{L}{l}\right)^d}
\end{align}
Applying the operator to this leads to two terms, we first take the limit for the $-\frac{1}{6}$ term
\begin{align}
&- \frac{1}{6} \beta_\deve S_{d-2}  \sum_{x=0}^2 \binom{2}{x} \frac{(-1)^{\dtwo-x}}{2^{\dtwo-1}} \frac{1}{(d+2)^2 } \binom{d-2}{\dtwo-x} \G{2+\twod}\\
&=\beta_\deve S_{d-2}  \frac{\G{2+\twod} \G{d-1}}{3 \; 2^{\dtwo -1} (d+2)^2 \G{1+\dtwo} \G{\dtwo}} \;.
\end{align}
For the second term we need to apply the derivative first
\begin{align}
& \lim_{l\to 0}  \frac{-l^{-(d+2)} \beta_\deve S_{d-2} }{24(d+1)(d+2)} \frac{d \; c_d \left( \frac{L}{l}\right)^d}{2^{\dtwo}} \nonumber \\
&\bigspace \sum_{x=0}^2 \binom{2}{x} \sum_{k=0}^{d-2} \binom{d-2}{k} \frac{(-1)^k \; L^{d+2}}{k+x+1+\dtwo}  \prod_{j=1}^{\dtwo} \frac{j+\dtwo}{j}\nonumber \\
& \bigspaced \mFm{\dtwo+1}{\twod(k+x+1)+1,\twod+2,\dots,3}{\twod(k+x+1)+2,\twod+1,\dots,2}{-c_d \left( \frac{L}{l}\right)^d} \;.
\end{align}
We can then take the limit, here only terms $k+x=\dtwo$ contribute 
\begin{align}
&\frac{- \beta_\deve S_{d-2} }{24(d+1)(d+2)} \frac{d }{2^{\dtwo}} \sum_{x=0}^2 \binom{2}{x} \binom{d-2}{\dtwo-x} \frac{(-1)^{\dtwo-x}}{d+1}  \prod_{j=1}^{\dtwo} \frac{j+\dtwo}{j}\nonumber \\
&\bigspaced \G{3+\twod} \prod_{j=1}^{\dtwo}\frac{j-1-\dtwo}{j+\dtwo} \frac{1}{c_d^{1+\twod}} \\
&=\frac{d \; \G{3+\twod}  \G{d-1} }{3 \; 2^{\dtwo+2} (d+1)^2 (d+2) \G{1 + \dtwo} \G{\dtwo} c_d^{1+\twod}}
\end{align}
We can then combine this with the term for $-\frac{1}{6}$ from above and find 
\begin{align}
\frac{1}{3} \beta_\deve S_{d-2} \frac{2^{\dtwo-3}\G{\frac{d-1}{2}} \G{\twod+1} }{(d+2)\sqrt{\pi} \G{1+\dtwo}c_d^{1+\twod}} \frac{ 5d+6 }{ d (d+1)}
\end{align}
for equation \eqref{eq:eveplusdef}.
The last integral is
\begin{align}
&\lim_{l\to 0} \beta_\deve l^{-(d+2)} S_{d-2} \times \nonumber \\
&\bigspace \frac{1}{12(d+1)(d+2)} \; l \frac{\partial}{\partial l}  \Od{d}  \int_{-L}^0 {\mathrm{d}}u  \!\! \int_{u}^0 \mathrm{d} v \, 
\bigg(\frac{v-u}{\sqrt{2}}\bigg)^{d-2} uv\;   e^{-l^{-d}V_{0\,d}} \\
&=\frac{\beta_\deve l^{-(d+2)} S_{d-2} }{3 \; 2^{\dtwo+1} (d+2)^2(d+1)} l \pd{}{l} \sum_{k=0}^{d-2} \binom{d-2}{k} \frac{(-1)^k}{k+2} L^{d+2} \nonumber\\
& \bigspace  \mFm{\dtwo+1}{\twod(k+2),\twod+1,\dots,2}{\twod(k+2)+1,\twod,\dots,1}{-c_d \left(\frac{L}{l}\right)^d}\\
&=\frac{\beta_\deve l^{-(d+2)} S_{d-2} \;d }{3 \; 2^{\dtwo+1} (d+2)^2(d+1)} \sum_{k=0}^{d-2} \binom{d-2}{k} \frac{(-1)^k}{k+2+\dtwo} L^{d+2}  c_d \left(\frac{L}{l}\right)^d \binom{d}{\dtwo}\nonumber\\
& \bigspace \mFm{\dtwo+1}{\twod(k+2)+1,\twod+2,\dots,3}{\twod(k+2)+2,\twod+1,\dots,2}{-c_d \left(\frac{L}{l}\right)^d}
\end{align}
We can then introduce $z$ and take the limit, in which only the term $k=\dtwo-1$ contributes 
\begin{align}
 &-  \frac{ 2^{\dtwo-3}}{3} \frac{d \beta_\deve  S_{d-2}}{(d+1)^2(d+2)^2 \; c_d^{1+\twod}} \frac{\G{3+\twod} \G{\frac{d-1}{2}}}{\G{\dtwo} \sqrt{\pi}}\;.
\end{align}

\subsection{\label{app:Rodd}Integrals in odd dimensions}
In odd dimensions we first calculate
\begin{align}
&\lim_{l\to 0} l^{-(d+2)}\beta_\dodd S_{d-2} \nonumber \\
& \bigspace \left(-\frac{1}{6} - \frac{1}{24(d+1)} l \pd{}{l} \right)  \frac{\Od{d}}{d-1}\intl{-L}{0} \md u \intl{u}{0} \md v \left( \frac{v-u}{\sqrt{2}} \right)^d e^{-l^{-d} V_{0\,d}} 
\end{align}
The integral with $\Od{d}$ acting on it has been calculated above to be \eqref{eq:oddintermediate}. We can then use this result and find
\begin{align}
&\lim_{l\to 0} l^{-(d+2)}\beta_\dodd \frac{S_{d-2}}{(d-1)(d+2)} \left(-\frac{1}{6} - \frac{1}{24(d+1)} l \pd{}{l} \right) \nonumber \\
&\ \ \ \ \ \ \ \ \ \ \  \sum_{k=0}^{d} \binom{d}{k} \frac{(-1)^k}{2^{\dtwo}} \frac{L^{d+2}}{(k+1)} \times \nonumber \\
&\bigspace \mFm{\frac{d+1}{2}+2}{1+\twod,\twod(k+1),\twod+1,\dots,2+\frac{1}{d}}{2+\twod,\twod(k+1)+1,\twod,\dots,1+\frac{1}{d}}{-c_d \left(\frac{L}{d}\right)^d}
\end{align}
For the term multiplied by $-\frac{1}{6}$ we can just use the limit \eqref{eq:oddlimitfin}
\begin{align}
-\frac{1}{3} \beta_\dodd S_{d-2}  \frac{2^{\dtwo} \G{\twod+1}}{d(d+1)(d-1) c_d^{1+\dtwo}} \;.
\end{align}
In the second term applying the derivative $l \pd{}{l}$ changes the arguments of the hypergeometric function
\begin{equation} 
\pd{}{z} \pFq{p}{q}{a_1, \dots ,a_p}{b_1, \dots ,b_q}{z} = \frac{ \prod_{j=1}^{p} a_j}{\prod_{j=1}^q b_q} \pFq{p}{q}{a_1+1, \dots ,a_p+1}{b_1+1, \dots ,b_q+1}{z} \;.
\end{equation}
Applying this we find
\begin{align}
&\frac{-l}{24(d+1)} \pd{}{l} \nonumber \\
&\bigspace \mFm{\frac{d+1}{2}+2}{1+\twod,\twod(k+1),\twod+1,\dots,2+\frac{1}{d}}{2+\twod,\twod(k+1)+1,\twod,\dots,1+\frac{1}{d}}{-c_d \left(\frac{L}{d}\right)^d} \\
=&\frac{-c_d d \left(\frac{L}{l}\right)^d}{24 (d+1)} \frac{ (d+2) (k+1)}{2 (d+1)(k+1+\dtwo)} \prod_{j=1}^{\frac{d+1}{2}} \frac{j+\dtwo}{j} \nonumber \\
&\bigspace \mFm{\frac{d+1}{2}+2}{2+\twod,\twod(k+1)+1,\twod+2,\dots,3+\frac{1}{d}}{3+\twod,\twod(k+1)+2,\twod+1,\dots,2+\frac{1}{d}}{-c_d \left(\frac{L}{d}\right)^d}\\
=& \frac{(d+2)(k+1) d \, \G{\frac{3}{2}+d}}{48 (d+1)^2 (k+1+\dtwo) \G{1+\dtwo} \G{\frac{3+d}{2}}} c_d \left(\frac{L}{l}\right)^d \nonumber \\
&\bigspace \mFm{\frac{d+1}{2}+2}{2+\twod,\twod(k+1)+1,\twod+2,\dots,3+\frac{1}{d}}{3+\twod,\twod(k+1)+2,\twod+1,\dots,2+\frac{1}{d}}{-c_d \left(\frac{L}{d}\right)^d} \;.
\end{align}
Putting this into the expression above and introducing $z$ we get
\begin{align}
&\lim_{z\to \infty} \sum_{k=0}^{d} \binom{d}{k} \frac{(-1)^k}{3 \,2^{\dtwo+4} } \frac{\G{\frac{3}{2}+d} }{\G{1+\dtwo} \G{\frac{3}{2}+\dtwo} } \frac{d \; \beta_d S_{d-2} }{(d+1)^2 (d-1) (k+1+\dtwo) c_d^{1+\twod}} \nonumber\\
& \ \ \ \ \ \ \  z^{2+\twod}  \mFm{\frac{d+1}{2}+2}{2+\twod,\twod(k+1)+1,\twod+2,\dots,3+\frac{1}{d}}{3+\twod,\twod(k+1)+2,\twod+1,\dots,2+\frac{1}{d}}{- z} \;.
\end{align}
To take the limit $z \to \infty$ we can use the same type of argument as above, first splitting the hypergeometric function using \eqref{eq:expident} and then examining the asymptotic behavior.
The only contribution can be from the part proportional to $ z^{2+\twod} \mFm{\frac{d+1}{2}+2}{2+\twod,\twod+2,\dots,3+\frac{1}{d}}{3+\twod,\twod+1,\dots,2+\frac{1}{d}}{- z}$. We expand it using equation \eqref{eq:limit} and obtain
\begin{align}
&\beta_\dodd S_{d-2} \sum_{k=0}^{d} \binom{d}{k} \frac{(-1)^{k+\frac{d+1}{2}}}{3 \;2^{\dtwo+4}} \frac{\G{\frac{3}{2}+d} \G{3+\twod} }{\G{1+\dtwo} \G{\frac{3}{2}+\dtwo} } \times  \nonumber \\
& \bigspaced \frac{d}{ (d+1)^2 (d-1) c_d^{1+\twod}} \frac{1}{\dtwo-k} \prod_{j=1}^\frac{d+1}{2} \frac{\dtwo+1-j}{j+\dtwo} \\
&=-\frac{2^{\dtwo-3}}{3} \frac{\G{3+\twod}}{(1+d)^3 (d-1)} \frac{d}{c_d^{1+\dtwo}} \beta_d S_{d-2} \;.
\end{align}
We can then sum this with the term above and find the result to \eqref{eq:oddminusdef}
\begin{align}
-\frac{1}{3} \beta_\dodd S_{d-2} \frac{2^\dtwo \G{\twod+1}}{(d+1)(d-1) c_d^{1+\twod}} \frac{5d+6}{4d(d+1)}
\end{align}
The next term we need to calculate is
\begin{align}
&\lim_{l\to 0} l^{-(d+2)}\beta_\dodd S_{d-2} \times \nonumber\\
& \ \ \ \ \ \left(-\frac{1}{6} - \frac{1}{24(d+1)} l \pd{}{l} \right) \Od{d}\intl{-L}{0} \md u \intl{u}{0} \md v \left( \frac{v-u}{\sqrt{2}} \right)^{d-2} \left( \frac{v+u}{\sqrt{2}} \right)^2 e^{-l^{-d} V_{0\,d}} 
\\
&=\lim_{l\to 0} l^{-(d+2)}\beta_\dodd S_{d-2} \times \nonumber \left(-\frac{1}{6} - \frac{1}{24(d+1)} l \pd{}{l} \right)\\
&\ \ \ \ \ \ \ \ \sum_{x=0}^2 \binom{2}{x} \sum_{k=0}^{d-2}\binom{d-2}{k} \frac{(-1)^k}{2^{\dtwo} }\frac{L^{-(d+2)}}{(k+x+1)(d+2)} \nonumber \\
&\ \ \ \ \ \ \ \ \ \ \ \ \ \ \ \mFm{\frac{d+1}{2}+2}{1+\twod,\twod(k+x+1),\twod+1,\dots,2+\frac{1}{d}}{2+\twod,\twod(k+x+1)+1,\twod,\dots,1+\frac{1}{d}}{-c_d \left(\frac{L}{l}\right)^d} \;.
\end{align}
For the $-\frac{1}{6}$ term  we introduce $z$ 
\begin{align}
&\lim_{z \to \infty} \beta_\dodd S_{d-2} \nonumber \sum_{x=0}^2 \binom{2}{x} \sum_{k=0}^{d-2}\binom{d-2}{k} \frac{(-1)^{k+1}}{3 \;2^{\dtwo+1} }\frac{z^{1+\twod}}{(k+x+1)(d+2) c_d^{1+\twod}} \\
& \bigspace \mFm{\frac{d+1}{2}+2}{1+\twod,\twod(k+x+1),\twod+1,\dots,2+\frac{1}{d}}{2+\twod,\twod(k+x+1)+1,\twod,\dots,1+\frac{1}{d}}{-z}
\end{align}
and then use identities \eqref{eq:expident},\eqref{eq:limit0} and \eqref{eq:limit} to take the limit. The result is
\begin{align}
-\frac{\beta_\dodd S_{d-2}}{d+2} \sum_{x=0}^2 \binom{2}{x} \sum_{k=0}^{d-2}\binom{d-2}{k} \frac{(-1)^{k+1}}{3 \;2^{\dtwo+1} } \frac{\G{2+\twod}}{(\dtwo-k-x)} \frac{\sqrt{\pi}}{\G{-\dtwo}\G{\frac{3+d}{2}}}\\
=\frac{\beta_\dodd S_{d-2} 2^{\dtwo} \G{\twod+1}}{3 d(d+1)(d-1)} \;.
\end{align}
For the second term, $- \frac{l}{24 (d+1)}\pd{}{l}$ we need to take the derivative before taking the limit. This leads to
\begin{align}
&\lim_{z \to \infty} \frac{\beta_\dodd S_{d-2}}{(d+1)^2} \frac{d \;\G{\frac{3}{2}+d}}{\sqrt{\pi} \G{2+d}} \nonumber \sum_{x=0}^2 \binom{2}{x} \sum_{k=0}^{d-2}\binom{d-2}{k} \frac{(-1)^{k+1}\; z^{2+\twod}\;2^{\dtwo-3} }{3 (k+x+1+\dtwo)c_d^{1+\twod}}\\
& \ \ \ \ \ \ \ \ \ \ \ \mFm{\frac{d+1}{2}+2}{2+\twod,\twod(k+x+1)+1,\twod+2,\dots,3+\frac{1}{d}}{3+\twod,\twod(k+x+1)+2,\twod+1,\dots,2+\frac{1}{d}}{-z}
\end{align}
where we already introduced $z$. The limit is then
\begin{align}
& \frac{\beta_\dodd S_{d-2}}{3 (d+1)^2} \frac{d \;2^{\dtwo-3}  \G{3+\twod}\G{1+\dtwo}}{\G{2+d} \G{-\dtwo}} \sum_{x=0}^2 \binom{2}{x} \sum_{k=0}^{d-2}\binom{d-2}{k} \frac{(-1)^{k+1}}{(\dtwo-k-x)}\\
 =& \beta_\dodd S_{d-2}\frac{2^{\dtwo-2} \G{2+\twod}}{3(d-1)(d+1)^2}
\end{align}
Summing these two terms then leads to the solution of \eqref{eq:oddplusdef}
\begin{align}
\beta_\dodd S_{d-2} \frac{2^{\dtwo} \G{1+\twod}}{3(d+1)(d-1)} \frac{5d+6}{4 d(d+1)}
\end{align}
The last integral for the odd case is
\begin{align}
&\lim_{l\to 0} \beta_\dodd l^{-(d+2)} S_{d-2} \times \nonumber \\
&\bigspace \frac{1}{12(d+1)(d+2)} \; l \frac{\partial}{\partial l}  \Od{d}  \int_{-L}^0 {\mathrm{d}}u  \!\! \int_{u}^0 \mathrm{d} v \, 
\bigg(\frac{v-u}{\sqrt{2}}\bigg)^{d-2} uv\;   e^{-l^{-d}V_{0\,d}} \\
=& \lim_{l\to 0} \beta_\dodd l^{-(d+2)} S_{d-2}
\frac{1}{12(d+1)(d+2)} \; l \frac{\partial}{\partial l}  \nonumber \\
&\bigspace \suml{k=0}{d-2} \binom{d-2}{k} \frac{(-1)^k}{2^{\dtwo -1}} \Od{d} \suml{n=0}{\infty} \frac{\left(-l^{-d}c_d \right)^n}{n!}  \frac{L^{d(n+1)+2}}{(\dtwo n+k+2)(d(n+1)+2)} \\
=& \lim_{l\to 0} \beta_\dodd l^{-(d+2)} S_{d-2} 
\frac{1}{12(d+1)(d+2)} \; l \frac{\partial}{\partial l}  \nonumber \\
&\bigspace \suml{k=0}{d-2} \binom{d-2}{k} \frac{(-1)^k}{2^{\dtwo -1}} \frac{L^{d+2}}{(k+2)(d+2)} \times  \nonumber \\
&\bigspace \mFm{\frac{d+1}{2}+2}{1+\twod, \twod(k+2),\twod +1, \dots, 2+\frac{1}{d}}{2+\twod, \twod(k+2)+1,\twod, \dots, 1+\frac{1}{d}}{-c_d \left( \frac{L}{l} \right)^d} \\
=& \lim_{l\to 0} \beta_\dodd l^{-(d+2)} S_{d-2} 
\frac{d}{12(d+1)^2(d+2)} c_d \left(\frac{L}{l}\right)^d \;   \nonumber \\
&\bigspace \suml{k=0}{d-2} \binom{d-2}{k} \frac{(-1)^k}{2^{\dtwo}} \frac{L^{d+2}}{(k+2+\dtwo)} \prod_{j=1}^{\frac{d+1}{2}} \frac{j+\dtwo}{j}\times  \nonumber \\
&\bigspace \mFm{\frac{d+1}{2}+2}{2+\twod, \twod(k+2)+1,\twod +2,  \dots, 3+\frac{1}{d}}{3+\twod, \twod(k+2)+2,\twod,  \dots, 2+\frac{1}{d}}{-c_d \left( \frac{L}{l} \right)^d} \\
=& \lim_{l\to 0} \beta_\dodd l^{-(d+2)} S_{d-2} 
\frac{d}{12(d+1)^2(d+2)} c_d \left(\frac{L}{l}\right)^d  \times \nonumber \\
&\bigspace \frac{\G{\frac{3}{2}+d}}{\G{1+\dtwo} \G{\frac{3+d}{2}}} \suml{k=0}{d-2} \binom{d-2}{k} \frac{(-1)^k}{2^{\dtwo}} \frac{L^{d+2}}{(k+2+\dtwo)} \times  \nonumber \\
& \bigspace \mFm{\frac{d+1}{2}+2}{2+\twod, \twod(k+2)+1,\twod +2, \dots, 3+\frac{1}{d}}{3+\twod, \twod(k+2)+2,\twod, \dots, 2+\frac{1}{d}}{-c_d \left( \frac{L}{l} \right)^d}
\end{align}
We can then introduce $z$ and take the limit. As above we split the function using \eqref{eq:expident}, see that the first term does not contribute and then take the limit for the second term, to find
\begin{align}
 &\beta_\dodd  S_{d-2} 
\frac{d}{12(d+1)^2(d+2)} \frac{\G{\frac{3}{2}+d}}{\G{1+\dtwo} \G{\frac{3+d}{2}}} \times \nonumber \\
& \bigspace \suml{k=0}{d-2} \binom{d-2}{k} \frac{(-1)^{k+1}}{2^{\dtwo}} \frac{\G{3+\twod}}{(\dtwo-1 - k)) c_d^{1+\twod}}  \prod_{j=1}^{\frac{d+1}{2}} \frac{j-1-\dtwo}{j+\dtwo}\\
&= -\frac{1}{3}\beta_\dodd  S_{d-2} \frac{2^{\dtwo-2} \G{1+\twod}}{(d-1)(d+1)^2 c_d^{1+\twod}}\;.
\end{align}

\end{appendix}
\end{document}